\documentclass[11pt,a4paper]{article}
\pdfoutput=1
\usepackage{jheppub}

\usepackage{multirow, graphicx,amssymb,url,mathrsfs,amsmath}
\usepackage{wrapfig,boxedminipage,setspace,subfigure,epsfig}
\usepackage{amsxtra,amstext,latexsym,dsfont,amsfonts}
\usepackage{color,eucal}
\usepackage[dvipsnames]{xcolor}
\usepackage{float}
\usepackage{slashed,comment}
\usepackage{kotex}

\renewcommand{\d}{\delta}

\newcommand{\dd}{\mathrm{d}}

\newcommand{\grad}{\nabla}

\title{Bound of diffusion constants from pole-skipping points: spontaneous symmetry breaking and magnetic field 
}

\author[a,b]{Hyun-Sik Jeong,}
\author[c]{Keun-Young Kim,}
\author[a,b]{and Ya-Wen Sun}

\emailAdd{hyunsik@ucas.ac.cn}
\emailAdd{fortoe@gist.ac.kr}
\emailAdd{yawen.sun@ucas.ac.cn}

\affiliation[a]{School of physics, University of Chinese Academy of Sciences, Beijing 100049, China}
\affiliation[b]{Kavli Institute for Theoretical Sciences, University of Chinese Academy of Sciences, Beijing 100049, China}
\affiliation[c]{School of Physics and Chemistry, Gwangju Institute of Science and Technology, \\
123 Cheomdan-gwagiro, Gwangju 61005, Korea}

\abstract{
We investigate the properties of pole-skipping of the sound channel in which the translational symmetry is broken explicitly or spontaneously.
For this purpose, we analyze, in detail, not only the holographic axion model, but also the magnetically charged black holes with two methods: the near-horizon analysis and quasi-normal mode computations. 
We find that the pole-skipping points are related with the chaotic properties, Lyapunov exponent ($\lambda_L$) and butterfly velocity ($v_B$), independently of the symmetry breaking patterns. 
We show that the diffusion constant ($D$) is bounded by $D \,\geqslant\, v_{B}^2/\lambda_{L} $, where $D$ is the energy diffusion (crystal diffusion) bound for explicit (spontaneous) symmetry breaking. We confirm that the lower bound is obtained by the pole-skipping analysis in the low temperature limit.

}

\begin{document}
\maketitle

\section{Introduction}


In recent years, the so-called pole-skipping~\cite{Grozdanov:2017ajz,Blake:2017ris,Blake:2018leo} has been one of interesting research topics in holography (or gauge/gravity duality)~\cite{Maldacena:1997re,Witten:1998qj,Gubser:1998bc} in that it not only can be used for the better understanding of the black hole horizon properties, but also help to study other interesting physics such as quantum chaos. 
Pole-skipping states that there is the special frequency and wave number in the momentum space ($\omega, k$), at which the Green's function could be ill-defined. This special point is called the pole-skipping point denoted by ($\omega_{*}, k_{*}$).

In holographic theories, pole-skipping might be understood from the black hole horizon properties~\cite{Blake:2018leo}.
According to the holographic dictionary, the Green's function of boundary correlators can be obtained by solving classical equation of motions which are second order differential equations. 
For general value of ($\omega, k$), the Green's function is determined by imposing the ingoing boundary condition at the horizon. However, at the pole-skipping point, it turns out that two independent ingoing solutions are available so that the Green's function is not uniquely defined.

\paragraph{Pole-skipping of the sound channel:}
The pole-skipping point is first discovered in the sound channel related to the retarded energy density two point functions, giving the connection with quantum chaos as 
\begin{align}\label{pspfirstint}
\begin{split}
\omega_{*} =  i \lambda_{L} \,, \qquad k_{*}  = i \frac{\lambda_{L}}{v_{B}} \,, 
\end{split}
\end{align}
where $\lambda_{L}=2\pi \,T$ is the Lyapunov exponent ($T$ is the temperature) and $v_{B}$ is a butterfly velocity, which characterizes the chaotic behavior~\cite{Grozdanov:2017ajz} via out-of-time-order correlators (OTOCs)~\cite{larkin1969quasiclassical, Kitaev-2014, Maldacena:2015waa}\footnote{OTOCs is one useful tool to study many-body quantum chaos: $\langle V_0(0) W_{\bf x}(t) V_0(0) W_{\bf x}(t) \rangle \sim 1 + e^{\lambda_{L} \left( t - x/v_{B} \right)}$. For OTOCs in the context of quantum many body systems and quantum mechanics we refer to \cite{Huh:2020joh, Hashimoto:2020xfr} and references therein.}.
Moreover, in addition to the non-uniqueness of ingoing solutions at the black hole horizon, it has also been shown that the pole-skipping points correspond to special points of momentum modes in the Einstein's equations near the black hole horizon: this method finding the pole-skipping point with near horizon properties is often referred to as the near-horizon analysis~\cite{Blake:2018leo}.

Inspired from the analysis of the energy density two point function, it has been realized that pole-skipping could also take place for other two point functions of operators dual to the bulk fields~\cite{Blake:2019otz,Grozdanov:2019uhi,Ceplak:2019ymw,Ceplak:2021efc}: for example, other components of the stress energy tensor and scalar, vector or fermionic (including the spin-3/2 case) operators.
However, unlike the energy density two point function, the pole-skipping in other cases turned out not to be related to chaos. See also recent developments of pole-skipping in \cite{Natsuume:2019sfp,Natsuume:2019xcy,Natsuume:2019vcv,Grozdanov:2018kkt,Grozdanov:2020koi,Li:2019bgc,Liu:2020yaf,Abbasi:2020ykq,Jansen:2020hfd,Wu:2019esr,Abbasi:2019rhy,Haehl:2018izb,Das:2019tga,Ramirez:2020qer,Liu:2020yaf,Ahn:2019rnq,Ahn:2020bks,Ahn:2020baf,Kim:2020url,Sil:2020jhr,Abbasi:2020xli}.


\paragraph{Pole-skipping with broken translational invariance:}
In addition to the connection with quantum chaos, the pole-skipping in the sound channel may also be related to other interesting physics when the translational symmetry is broken: the universal diffusion bound~\cite{Blake:2018leo}.

Holography with broken translational invariance has provided insights for strongly coupled systems. 
The study of universal diffusion bound would be a representative example, which is inspired by the Kovtun-Son-Starinets (KSS) bound~\cite{Kovtun:2004de}\footnote{This KSS bound may help to understand the universal features in the transport properties of strongly coupled systems in that it appeared to hold with realistic experimental data~\cite{Schafer:2009dj,Cremonini:2011iq,Luzum:2008cw,Nagle:2011uz,Shen:2011eg}, however it turns out that this bound can be easily violated by considering translational symmetry breaking~\cite{Alberte:2016xja,Hartnoll:2016tri,Burikham:2016roo,Rebhan:2011vd}. 
}. 
In holographic theories, it turned out that when the translational invariance is broken the diffusion constant ($D$) may have the following universal lower bounds\footnote{This bound is included to the case of the IR geometry is AdS$_{2}\times R^2$. There could be a pre-factor depending on the IR geometry or the details of the deformation away from the fixed point~\cite{Blake:2016jnn}. In this paper, for simplicity, we focus on the case of the AdS$_{2}\times R^2$ IR geometry, in other words, the diffusion bound would be the AdS$_{2}$ diffusion bound. The translational invariant system at finite chemical potential~\cite{Blake:2016jnn,Blake:2017qgd} also gives the energy diffusion and if the IR geometry is AdS$_2$, the same argument is valid. This bound between transport properties and chaos was first proposed in the holographic models. It has also been observed in condensed matter theories \cite{Zhang:2016ofh,Aleiner:2016eni,Swingle:2016jdj,Patel:2016wdy}.}
\begin{align}\label{eq12first}
\begin{split}
D \,\geqslant\, v_{B}^2/\lambda_{L} \,,
\end{split}
\end{align}
where $D$ depends on the the symmetry breaking patterns, i.e., it corresponds to the energy diffusion constant ($D_{E}$) for the explicit symmetry breaking (EXB)~\cite{Blake:2016sud,Blake:2017qgd,Blake:2016jnn,Kim:2017dgz,Ahn:2017kvc} and it is the crystal diffusion constant ($D_{\phi}$) for the spontaneous symmetry breaking (SSB)~\cite{Baggioli:2020ljz,Baggioli:2017ojd}. 
Here the bound (the equality) is approached when the temperature is very small compared to any other scales in given models.

The relation between the pole-skipping and the diffusion bound is as follows.
In the hydrodynamic regime (small $\omega$, small $k$), the hydrodynamic dispersion relation with the broken translational invariance could be the diffusion mode as $\omega\,=\,-i\,D \,k^2$ and, in general, the pole-skipping point is not supposed to be related to this diffusion mode.

However, when the translational symmetry is broken explicitly (i.e., the diffusion process is governed by the energy diffusion constant, $D_{E}$) it has been shown that the pole-skipping point could pass through the energy diffusion mode at low temperature, and produce the lower bound for the energy diffusion constant (i.e., $\omega_{*}\,=\,-i\,D_{E} \,k_{*}^2$ gives the lower bound $D_{E} = v_{B}^2 \,/ \lambda_{L}$)~\cite{Blake:2018leo}.

\paragraph{Motivations of this paper:}
Based on several developments above, we may summarize main consequences from the pole-skipping analysis of the sound channel as
\begin{itemize}
\item{\textbf{Universal features of Einstein's equation:} \\ From the near-horizon analysis, the pole-skipping point seems to be universal in that its form \eqref{pspfirstint} does not depend on the matter field perturbations, so the pole-skipping phenomenon may help to understand the hidden universal structure of Einstein equations.}
\item{\textbf{An independent method to compute the butterfly velocity:} \\
Pole-skipping can be considered as an independent method to compute quantum chaos quantities in holography. For instance, one can compute $v_{B}$ from the near-horizon analysis, which is consistent with the shock wave analysis result.} 
\item{\textbf{A new method to capture the diffusion bound:} \\
Pole-skipping can capture the universal diffusion bound which is mostly investigated with the IR analysis. Thus the pole-skipping analysis may also help to investigate the universal relations in strongly coupled systems.}
\end{itemize}
In spite of these interesting advantages, most of holographic studies focused on the case with the EXB. 
Therefore, the pole-skipping analysis with the SSB is still missing, and we will fill this gap in this paper.

Moreover, even for the EXB case, these features are only checked when the translational symmetry is broken by an axion field~\cite{Blake:2018leo}. In holography, when the magnetic fields ($H$) are turned on, we can also study the EXB without axion fields  so that the study of the pole-skipping with the EXB may also require further investigation.

To our knowledge, this approach has not been investigated yet.
Our goal here is to extend above three big items in three aspects. 
i) to study the pole-skipping with different symmetry breaking patterns: EXB vs SSB; 
ii) to understand the pole-skipping with different sources of the EXB: axion charge vs magnetic fields; 
iii) to extend the formalism in \cite{Blake:2018leo} to the case where $g_{tt}\neq g_{rr}^{-1}$ and $g_{xx} \neq g_{yy}$.

Considering symmetry breaking patterns, we will also perform quasi-normal mode computations as well as the near-horizon analysis for the following reasons: 
(i) to confirm that the pole-skipping point indeed passes through dispersion relations of the system. Note that the near-horizon analysis itself may not be complete to show that the special point is one of poles;
(ii) If (i) is checked, we may see which dispersion relation is related to the pole-skipping point. For instance, $\omega = v_{B} \, k$ is a mathematically possible candidate.

The second item (ii) is related to the study of the diffusion bound from the pole-skipping.
As we mentioned, the pole-skipping analysis with SSB is still missing. 
For the SSB, it might be more interesting because, unlike the EXB case, we have two dispersion relations, i.e. the longitudinal sound modes and the crystal diffusion mode.

This paper is organized as follows. 
In section 2, we introduce the Einstein-Maxwell-Dilaton with Axion model. Depending on the couplings, we show how this model can describe the EXB or the SSB of translational invariance. Then using the near-horizon analysis, we study the pole-skipping point of the sound channel in such models.
In section 3 using axion models, we show the (energy/crystal) diffusion bound can be obtained from the pole-skipping points with different symmetry breaking patterns.
In section 4 we study the magnetic field effect on the pole-skipping in the case of EXB of translational invariance.
Section 5 is devoted to conclusions.

\section{Near-horizon analysis and pole-skipping}

In this section, we will introduce the Einstein-Maxwell-Dilaton with Axion model (EMD-Axion model) and show that this model can describe the EXB or the SSB of  translational invariance depending on the couplings. Then we will study the pole-skipping of the sound channel in these models using the near horizon analysis.

\subsection{Einstein equations and pole-skipping}
We consider the following (3+1) dimensional Einstein gravity
\begin{equation}
\begin{split}
S = \int \dd^4x \sqrt{-g} \,\left( R \,+\, 6 \,+\, \mathcal{L}_{M} \right) \,, \label{GENAC}
\end{split}
\end{equation}
where $\mathcal{L}_{M}$ is the general matter Lagrangian and we set units such that the gravitational constant $16 \pi G=1$, and the AdS radius $L=1$.
Moreover, we write the background metric ansatz as 
\begin{equation}\label{METANS}
\begin{split}
\dd s^2 =  -D(r)\, \dd t^2 +  B(r) \, \dd r^2  + C_{1}(r) \dd x^2 + C_{2}(r) \dd y^2 \,.
\end{split}
\end{equation}
Note that this general metric becomes the one in \cite{Blake:2018leo} when
\begin{equation}\label{RICHARDIDENTITY}
\begin{split}
D(r) = \frac{1}{B(r)} \,, \quad C_{1}(r) = C_{2}(r).
\end{split}
\end{equation}
Thus our analysis is applicable for the case of general black holes, for instance, anisotropic Q-lattice models in which the metric can be taken as $D(r) \neq \frac{1}{B(r)}$ or $C_{1}(r) \neq C_{2}(r)$. 

From metric \eqref{METANS}, the Hawking temperature is given by
\begin{equation}\label{HT}
   T = \frac{1}{4\pi } \left. \frac{|D'|}{\sqrt{D B}}\right|_{r_{h}} \,,
\end{equation}
where $r_{h}$ is the black hole horizon.

For the pole-skipping analysis, it is convenient to introduce the incoming Eddington-Finkelstein coordinate $v$
\begin{equation}\label{}
    v = t + r_{*}\,, \quad \dd r_{*} = \sqrt{\frac{B(r)}{D(r)}} \, \dd r \,,
\end{equation}
in terms of which the metric becomes 
\begin{equation}\label{METICANSAY}
\begin{split}
\dd s^2 =  -D(r)\, \dd v^2 +  2 \sqrt{D(r) B(r)} \, \dd v \dd r  + C_{1}(r) \, \dd x^2 + C_{2}(r) \, \dd y^2 \,.
\end{split}
\end{equation}
Based on this background metric, we consider the fluctuations 
\begin{equation}\label{MFF}
\begin{split}
\delta g_{\mu\nu} (r, v, x, y) = \delta \bar{g}_{\mu\nu}(r) \, e^{-i\omega v + i k x} \,, \quad \delta \Phi (r, v, x, y) = \delta \bar{\Phi}(r) \, e^{-i\omega v + i k x} \,,
\end{split}
\end{equation}
where $\delta \Phi$ represents general matter field perturbations that couple to gravitational perturbations.
The pole-skipping phenomena is related to the near-horizon properties of Einstein equations. 
In particular, the constraint from the $vv$ component of Einstein equation could vanish at the pole-skipping points related to the quantum chaos, then leading to the existence of an extra independent ingoing solution, which makes the Green's function multiple-valued at the pole-skipping point~\cite{Blake:2018leo}.

In order to understand this, we expand the fluctuations near the horizon as
\begin{align}\label{}
\begin{split}
\delta\bar{g}_{\mu\nu}(r) &= \delta\bar{g}_{\mu\nu}^{(0)} \,+\, \delta\bar{g}_{\mu\nu}^{(1)}  (r-r_{h}) \,+\, \dots \,, \\
\delta\bar{\Phi}(r) &= \delta\bar{\Phi}^{(0)} \,+\, \delta\bar{\Phi}^{(1)}  (r-r_{h}) \,+\, \dots \,.
\end{split}
\end{align}
Then the $vv$ component of Einstein equations reads
\begin{align}\label{PSE}
\begin{split}
\left(-i \omega \, \frac{(C_{1}^{H} C_{2}^{H})^{'}}{2\, C_{2}^{H} \sqrt{D^{H}\,B^{H}}} + k^2\right) \delta\bar{g}_{vv}^{(0)}  - i (2\pi T &+ i \omega)\left[ \omega\left(\delta\bar{g}_{xx}^{(0)} +\frac{C_{1}^{H}}{C_{2}^{H}}\delta\bar{g}_{yy}^{(0)}\right) + 2 k \delta\bar{g}_{vx}^{(0)}  \right]  \\  
= & -2 C_{1}^{H} \left(  \frac{T_{vr}^{H}}{\sqrt{D^{H} B^{H}}} \, \delta \bar{g}_{vv}^{(0)} - \delta T_{vv}^{H}   \right)  \,,
\end{split}
\end{align}
where the superscript ($H$) denotes quantities evaluated at the horizon, $T_{\mu\nu}$ is the background stress-energy tensor and $\delta T_{\mu\nu}$ represents the perturbed one by matter fields.

We note two things here. First, the gravitational fluctuations in this equation correspond to the sound channel~\cite{Kovtun:2005ev}
\begin{align}\label{GSM}
\begin{split}
\{\delta g_{vv} \,,\,\,  \delta g_{vx} \,,\,\,  \delta g_{xx} \,,\,\,  \delta g_{yy} \}\,.
\end{split}
\end{align}
Second, the following identity, from the right hand side of \eqref{PSE}, is checked for the simple Ansatz \eqref{RICHARDIDENTITY} \cite{Blake:2018leo},
\begin{align}\label{IDENTI}
\begin{split}
  \frac{T_{vr}^{H}}{\sqrt{D^{H} B^{H}}} \, \delta \bar{g}_{vv}^{(0)} - \delta T_{vv}^{H} = 0 \,.
\end{split}
\end{align}
We will shortly show that \eqref{IDENTI} still holds for general metric \eqref{METICANSAY} in the next subsection.

Using \eqref{IDENTI}, for generic value of $\omega$, the equation \eqref{PSE} imposes the constraints between the horizon values of metric components: 
$\delta \bar{g}_{vv}^{(0)} \,, \delta \bar{g}_{vx}^{(0)} \,, \delta \bar{g}_{xx}^{(0)}$ and $\delta \bar{g}_{yy}^{(0)}$.
However, when the frequency takes $\omega = \omega_{*} = i 2\pi T$\,,  equation \eqref{PSE} simplifies as
\begin{align}\label{TEQTEQ}
\begin{split}
\left(  2\pi T \, \frac{(C_{1}^{H} C_{2}^{H})^{'}}{2\, C_{2}^{H} \sqrt{D^{H}\,B^{H}}} + k^2\right) \delta\bar{g}_{vv}^{(0)} = 0 \,.
\end{split}
\end{align}
%
For the generic wave number $k$, this equation gives $\delta\bar{g}_{vv}^{(0)} = 0$, however when 
\begin{align}\label{}
\begin{split}
k^2 = k_{*}^2 = -2\pi T \, \frac{(C_{1}^{H} C_{2}^{H})^{'}}{2\, C_{2}^{H} \sqrt{D^{H}\,B^{H}}} \,,
\end{split}
\end{align}
equation \eqref{TEQTEQ} is automatically satisfied.

Therefore, assuming the non-trivial identity \eqref{IDENTI}, equation \eqref{PSE} at the following special point \eqref{PSP} is identically satisfied and we cannot impose constraints between the horizon metric components, leading to pole-skipping in the two-point functions.
\begin{align}\label{PSP}
\begin{split}
\omega = \omega_{*} =  i \lambda_{L} \,, \qquad k = k_{*}  = i \frac{\lambda_{L}}{v_{B}} \,, 
\end{split}
\end{align}
where $\lambda_{L} = 2\pi T$ is the Lyapunov exponent and the butterfly velocity $v_{B}$ is  
\begin{align}\label{VTFOMU}
\begin{split}
\frac{1}{v_{B}} = \sqrt{ \frac{(C_{1}^{H} C_{2}^{H})^{'}}{4 \pi T\, C_{2}^{H} \sqrt{D^{H}\,B^{H}}} } \,,
\end{split}
\end{align}
which is consistent with the one by shock-wave analysis~\cite{Ahn:2017kvc}.

\subsection{Einstein-Maxwell-Dilaton with Axion model with symmetry breaking patterns}

In order to check the identity \eqref{IDENTI}, we need to specify the form of matter Lagrangian ($\mathcal{L}_M$) in \eqref{GENAC}. 
In this paper, we choose the following Einstein-Maxwell-Dilaton with Axion theories
\begin{equation}\label{GENCLASS}
\mathcal{L}_M =  - \frac{1}{2}\left(\partial\phi\right)^2 + V(\phi) -\frac{Z(\phi)}{4}F^2 - Y(\phi) \, W(X),
\end{equation}
where
\begin{align}\label{}
\begin{split}
X := \frac{1}{2}\sum_{i=1}^{2}\left(\partial\varphi_i\right)^2\,.
\end{split}
\end{align}
This model is composed of three matter fields 
\begin{align}
\label{matter}
\phi = \phi(r), && \varphi_i = m \, x^i,&& A = A_v(r) \dd v -\frac{H}{2} y \,\dd x \,+\, \frac{H}{2} x \, \dd y,
\end{align}
where $\phi$ is the dilaton field originally introduced to avoid a finite entropy at zero temperature and $\varphi_{i}$ is the axion field added to break the translational invariance so that momentum could be relaxed, and $H$ is an external magnetic field.

There are three main reasons to consider \eqref{GENCLASS}.
First, it contains a large class of holographic models. For instance, \eqref{GENCLASS} becomes the one in \cite{Blake:2018leo} when
\begin{equation}\label{}
W(X) = X \,, \quad H = 0 \,.
\end{equation}
Thus we could study pole-skipping in more general holographic models using \eqref{GENCLASS}.

Second, with the choice of $W(X)$ in \eqref{GENCLASS}, we could study both the explicitly broken translational invariance (EXB) and the spontaneously broken translational invariance (SSB), so we could investigate pole-skipping with different types of symmetry breaking patterns.  

Third, we may further study the relationship between pole-skipping and the EXB of translational invariance. 
There are two simple ways to study the EXB of translational invariance in holography by adding: i) axion charge ($m$); ii) magnetic fields ($H$). 
Unlike the case with axion charge~\cite{Blake:2018leo},  pole-skipping with magnetic fields has not been fully studied yet, in particular, from the perspective of the quasi-normal mode spectrum.

\subsubsection{Translational symmetry breaking in holography}
In the following we explain the method that we use in this work to obtain (EXB, SSB) in holography.  
From \eqref{GENCLASS}, the equation of motion for the axion field becomes 
\begin{equation}\label{}
\grad_{\mu}\left[ Y(\phi) \, W'(X) \,\grad^{\mu}\varphi_i \right] =0 \,,
\end{equation}
and, in general, we can expand the couplings and the axion field at the AdS boundary ($r\rightarrow0$)
\begin{equation}\label{}
Y(\phi) \, W'(X) \,\sim\, r^{n} \,, \qquad \varphi_{i} \,\sim\,  \varphi_{i}^{(0)}(x^{i}) \, r^{0} \, + \, \varphi_{i}^{(\Delta)}(x^{i}) \, r^{\Delta} \, + \dots \,.
\end{equation}
Then, one can find the following relation
\begin{equation}\label{}
\Delta \,=\, 3-n  \,,
\end{equation}
i.e., the power $n$ of couplings determines the axion field's boundary behavior ($\Delta$). 
In other words, tuning $n$ could make $\varphi_{i}^{(0)}(x^{i})$ either the leading order or the sub-leading order:
\begin{align}\label{BOUNDARYEXP}
\begin{cases}
n  < 3 :     \quad \Delta > 0  \quad  \, \rightarrow  \quad\, \varphi_{i}^{(0)}(x^{i}) \,\,  \text{is \,the\, \textit{leading} \,order}\,, \\
n  > 3 :     \quad \Delta < 0  \quad  \, \rightarrow \quad\, \varphi_{i}^{(0)}(x^{i}) \,\,  \text{is \,the\, \textit{sub-leading}\, order}\,.
\end{cases}
\end{align}

According to the holographic dictionary, the leading order is interpreted as the source and the sub-leading order is associated with the expectation value of the operator conjugate to the source. Thus, with \eqref{BOUNDARYEXP}, we can consider SSB by setting 
\begin{align}\label{CONSSB}
n  > 3 \,, \qquad  \varphi_{i}^{(\Delta)}(x^{i}) = 0 \,,
\end{align}
i.e., the vacuum expectation value (the sub-leading order)  $\varphi_{i}^{(0)}(x^{i}) = m \, x^i$, is induced without the source (the leading order) $\varphi_{i}^{(\Delta)}(x^{i})$.
Therefore, we can generally write conditions for each symmetry breaking patterns as
\begin{align}\label{}
\begin{cases}
n < 3 :     \quad\rightarrow\quad   \Delta > 0 \qquad\qquad\qquad\qquad\qquad\qquad\, (\text{EXB}) \,, \\
n > 3 :     \quad\rightarrow\quad   \Delta < 0 \qquad\&\qquad \varphi_{i}^{(\Delta)}(x^{i}) = 0 \qquad (\text{SSB}) \,.
\end{cases}
\end{align}
Thus previous holographic studies of SSB~\cite{Ammon:2019wci,Ammon:2019apj,Alberte:2017oqx,Alberte:2017cch,Andrade:2017cnc,Amoretti:2017frz,Amoretti:2018tzw,Amoretti:2019cef,Amoretti:2019kuf,Baggioli:2021xuv} might be explained within this general analysis. 
\paragraph{Example 1:} For instance, when the holographic models do not contain the dilaton field  
\begin{equation}\label{}
Y(\phi) = 1 \,,\,\,  W(X) = X^{N}  \quad\rightarrow\quad Y(\phi) \, W'(X) \,\sim\, r^{2(N-1)}\,,
\end{equation}
where $X \sim r^{2}$ near the boundary and $n$ is $2(N-1)$. Thus $\Delta=5-2N$, and the symmetry breaking patterns can be determined as
\begin{align}\label{MATTEOMODEL}
\begin{cases}
N< 5/2 :     \quad\rightarrow\quad   \Delta > 0 \qquad (\text{EXB}) \,, \\
N> 5/2 :     \quad\rightarrow\quad   \Delta < 0 \qquad (\text{SSB}) \,.
\end{cases}
\end{align}
Note that holographic models in previous study of pole-skipping \cite{Blake:2018leo} corresponds to the case of $N=1$ which belongs to EXB.

\paragraph{Example 2:} It is also applicable when the dilaton field is added. For instance, 
\begin{equation}\label{}
Y(\phi) = \phi^{2} \,,\,\, W(X) = X  \quad\rightarrow \quad  Y(\phi) \, W'(X) \,\sim\, \phi^2 \,.
\end{equation}
Then with the asymptotic behavior of dilaton field,
\begin{equation}\label{}
 \phi \sim \lambda\, r + \mathcal{O}\, r^2
\end{equation}
the symmetry breaking pattern can be classified as
\begin{align}\label{}
\begin{cases}
\lambda \neq 0 :    \,\,  Y(\phi) \, W'(X)  \,\,\sim\,\, r^{2} \quad\rightarrow\quad  \Delta = 3-2 = +1 \quad (\text{EXB}) \,, \\
\lambda = 0 :        \,\,   Y(\phi) \, W'(X)  \,\,\sim\,\, r^{4} \quad\rightarrow\quad  \Delta = 3-4 = -1 \quad (\text{SSB}) \,.
\end{cases}
\end{align}

\subsubsection{Stress-energy tensor}
Now, let us check the identity \eqref{IDENTI} for \eqref{GENCLASS}. One can consider this subsection as the generalization of appendix A in \cite{Blake:2018leo} in three aspects: i) general metric; ii) all symmetry breaking patterns; iii) finite magnetic fields.

The stress-energy tensor for \eqref{GENCLASS} is given by
\begin{equation}
\begin{aligned}
\label{GENCLASSstress}
T_{\mu\nu}&=\frac{1}{2}\mathcal{L}_M g_{\mu\nu}-\frac{\partial\mathcal{L}_M}{\partial g^{\mu\nu}}\\
&=\frac{1}{2}\mathcal{L}_M g_{\mu\nu}+\frac{1}{2}\partial_\mu\phi\partial_\nu\phi+\frac{Y(\phi)}{2} W'(X) \sum_{i=1}^{2}\partial_\mu\varphi_i\partial_\nu\varphi_i-\frac{Z(\phi)}{2}F_{\mu\alpha}g^{\alpha\beta}F_{\beta\nu}\,,
\end{aligned}
\end{equation}
from which we obtain the component $T_{vr}$ in \eqref{IDENTI} with the metric \eqref{METICANSAY} as 
\begin{equation}
\label{GENCLASSstressRV}
T_{vr}=\frac{1}{2}\mathcal{L}_M \sqrt{D B}-  \frac{Z(\phi)}{2\,\sqrt{D B}}F_{vr}^2.
\end{equation}

For $\delta T_{vv}$ in \eqref{IDENTI}, we need to fluctuate the stress tensor \eqref{GENCLASSstress} with respect to fields
\begin{equation}
\begin{aligned}
&\delta T_{\mu\nu}=\frac{1}{2}\mathcal{L}_M\delta g_{\mu\nu}+\frac{Z(\phi)}{2}F_{\mu\alpha}g^{\alpha\gamma}g^{\delta\beta}F_{\beta\nu}\delta g_{\gamma\delta}+\frac{1}{2}g_{\mu\nu}\left[\frac{\delta \mathcal{L}_M}{\delta g^{\alpha\beta}}\delta g^{\alpha\beta}+\frac{\delta \mathcal{L}_M}{\delta\psi_i}\delta\psi_i\right]\\
&-\frac{Z(\phi)}{2}g^{\alpha\beta}\left[F_{\mu\alpha}\left(\partial_\beta\delta A_\nu-\partial_\nu\delta A_\beta\right)+F_{\nu\alpha}\left(\partial_\beta\delta A_\mu-\partial_\mu \delta A_\beta\right)\right]-\frac{Z'(\phi)}{2}F_{\mu\alpha}g^{\alpha\beta}F_{\beta\nu}\delta\phi \\ 
&+\frac{1}{2}\left[\partial_\mu\phi\partial_\nu\delta\phi+\partial_\mu\delta\phi\partial_\nu\phi+Y(\phi) W'(X) \sum_{i=1}^{2}\left(\partial_\mu\varphi_i\partial_\nu\delta\varphi_i+\partial_\nu\varphi_i\partial_\mu\delta\varphi_i\right) \right]  \\
&+\frac{\delta\phi}{2} \sum_{i=1}^{2}\partial_\mu\varphi_i\partial_\nu\varphi_i \left[ Y'(\phi) W'(X)  + Y(\phi) W''(X) \right] \,,
\end{aligned}
\end{equation}
where $\psi_i=\{A_\mu,\phi,\varphi_i\}$ denotes the matter fields.
Note that the very last term $W''(X)$ is a new term that did not appear in \cite{Blake:2018leo}.

Then, $\delta T_{vv}$ reads with the metric \eqref{METICANSAY} as
\begin{eqnarray}\label{GENCLASSstressRV2}
\delta T_{vv} &=& \left(\frac{1}{2}\mathcal{L}_M-\frac{Z(\phi)}{2 DB}F_{vr}^2\right)\delta g_{vv}
+ g_{vv}\bigg[-\frac{Z(\phi)}{2(DB)^2}g_{vv}F_{vr}^2\delta g_{rr}+ \frac{Z(\phi)}{(DB)^{3/2}} F_{vr}^2 \delta g_{vr}  \nonumber \\ &+& \frac{1}{2}\frac{\delta \mathcal{L}_M}{\delta g^{\alpha\beta}}\delta g^{\alpha\beta} + \frac{1}{2}\frac{\delta \mathcal{L}_M}{\delta\psi_i}\delta\psi_i + \frac{Z(\phi)}{DB} F_{vr}\left(\partial_r\delta A_v-\partial_v\delta A_r\right)-\frac{Z'(\phi)}{2 DB}F_{vr}^2\delta\phi\bigg],
\end{eqnarray}
where we have used $g^{rv} = \frac{1}{\sqrt{DB}}, \, g^{rr} = - g_{vv} \frac{1}{DB}$.

Evaluating \eqref{GENCLASSstressRV} with \eqref{GENCLASSstressRV2} at the horizon, with $g_{vv}(r_h) = 0$\footnote{We also need to assume the quantity in square brackets in \eqref{GENCLASSstressRV2} is regular at the horizon \cite{Blake:2018leo}.}, we can see that 
\begin{align}\label{IDENTI2}
\begin{split}
  \frac{T_{vr}^{H}}{\sqrt{D^{H} B^{H}}} \, \delta \bar{g}_{vv}^{(0)} - \delta T_{vv}^{H} = 0 
\end{split}
\end{align}
which means \eqref{IDENTI} still holds for holographic models \eqref{GENCLASS} and pole-skipping point \eqref{PSP} is universal in that: i) the metric is general 
including anisotropy; ii) the magnetic field is included; iii) it does not depend on the symmetry breaking pattern.

\section{Pole-skipping and symmetry breaking patterns}

In this section, we study pole-skipping with symmetry breaking patterns.
In particular, we focus on the following axion models\footnote{For the recent development of this model, see \cite{Baggioli:2021xuv} and references therein.} 
\begin{align}\label{genaxionmodel}
\begin{split}
\mathcal{L}_{M} = -\left[\frac{1}{2}  \sum_{i=1}^{2}(\partial\varphi_{i})^2 \right]^{N} \,, \quad \varphi_{i} = m x^{i} \,,
\end{split}
\end{align}
which corresponds to \eqref{GENCLASS} with
\begin{align}\label{}
\begin{split}
\phi = 0 \,,\quad  V(\phi) = 0  \,,\quad  Z(\phi) = 0 \,,\quad Y(\phi) = 1 \,,\quad W(X) = X^{N} \,,
\end{split}
\end{align}
and
\begin{align}\label{}
\begin{split}
A_{v} = 0 \,,\quad H = 0\,.
\end{split}
\end{align}

This model \eqref{genaxionmodel} allows an analytic background solution of 
\begin{equation}\label{}
\begin{split}
 D(r)\,= \frac{1}{B(r)} = r^2 - \frac{m_{0}}{r} \,+ \, \frac{m^{2N}}{2\,(2N-3)\,r^{2N-2}} \,, \quad C_{1}(r) = C_{2}(r) =  r^2 \,,
\end{split}
\end{equation}
where $m_{0}$ is determined by the condition $D(r_{h})=0$:
\begin{align}\label{}
m_{0} = r_{h}^3\left( 1 +  \frac{m^{2N}}{(4N-6)\, r_{h}^{2N}} \right)\,, \qquad r_{h} = \text{ the radius of horizon.}
\end{align}

In this background, the butterfly velocity \eqref{VTFOMU} reads as 
\begin{align}\label{VTFOMU22}
v_{B}^2 = \frac{\pi \, T}{r_{h}} \,.
\end{align}
The Hawking temperature \eqref{HT} and other thermodynamic quantities~\cite{Ammon:2020xyv} are
\begin{align}\label{THMERRESULTSGEN}
\begin{split}
 T &\,=\, \frac{1}{4\pi} \left( 3\,r_{h} \,-\, \frac{m^{2N}}{2\,r_{h}^{2N-1}}  \right)  \,, \qquad\,\,   s \,=\, 4\pi r_{h}^2  \,, \\ 
 \epsilon &\,=\, 2 \,r_{h}^3  - \frac{2 m^{2N}}{6-4N}\,r_{h}^{3-2N}  \,, \qquad\quad   P \,= \, r_{h}^3 + \frac{(2N-1)\,m^{2N}}{6-4N} \, r_{h}^{3-2N}  \,,
\end{split}
\end{align}
where $(s, \epsilon, P)$ are the entropy density, energy density, and the pressure respectively. Here we introduce the energy diffusion constant ($D_{E}$) for later use as
\begin{align}\label{nsgodxcew}
D_{E} := \frac{\kappa}{c_{\rho}} \,, \qquad  \kappa = \frac{16 \pi^2 T}{N} \left(\frac{r_{h}}{m}\right)^{2N} \,, \quad c_{\rho} := T\frac{\partial s}{\partial T}\,, 
\end{align}
where $\kappa$ is the thermal conductivity~\cite{Baggioli:2016pia} and $c_{\rho}$ is the specific heat.

In this axion model \eqref{genaxionmodel}, the sound channel with the matter fluctuation $\delta \Phi$ \eqref{MFF} is as follows 
%
\begin{align}\label{PW2PW2PW2}
\begin{split}
\{\delta g_{vv}, \,\delta g_{vx}, \,\delta g_{xx}, \, \delta g_{yy},\, \delta \varphi_{x}\} \,.
\end{split}
\end{align} 

\subsection{Explicit symmetry breaking: a review}

Let us first study the explicit symmetry breaking case (EXB)  \eqref{MATTEOMODEL}.
In particular, we will mainly review \cite{Blake:2018leo} in a self-contained manner.

\begin{figure}[]
\centering
     {\includegraphics[width=7.2cm]{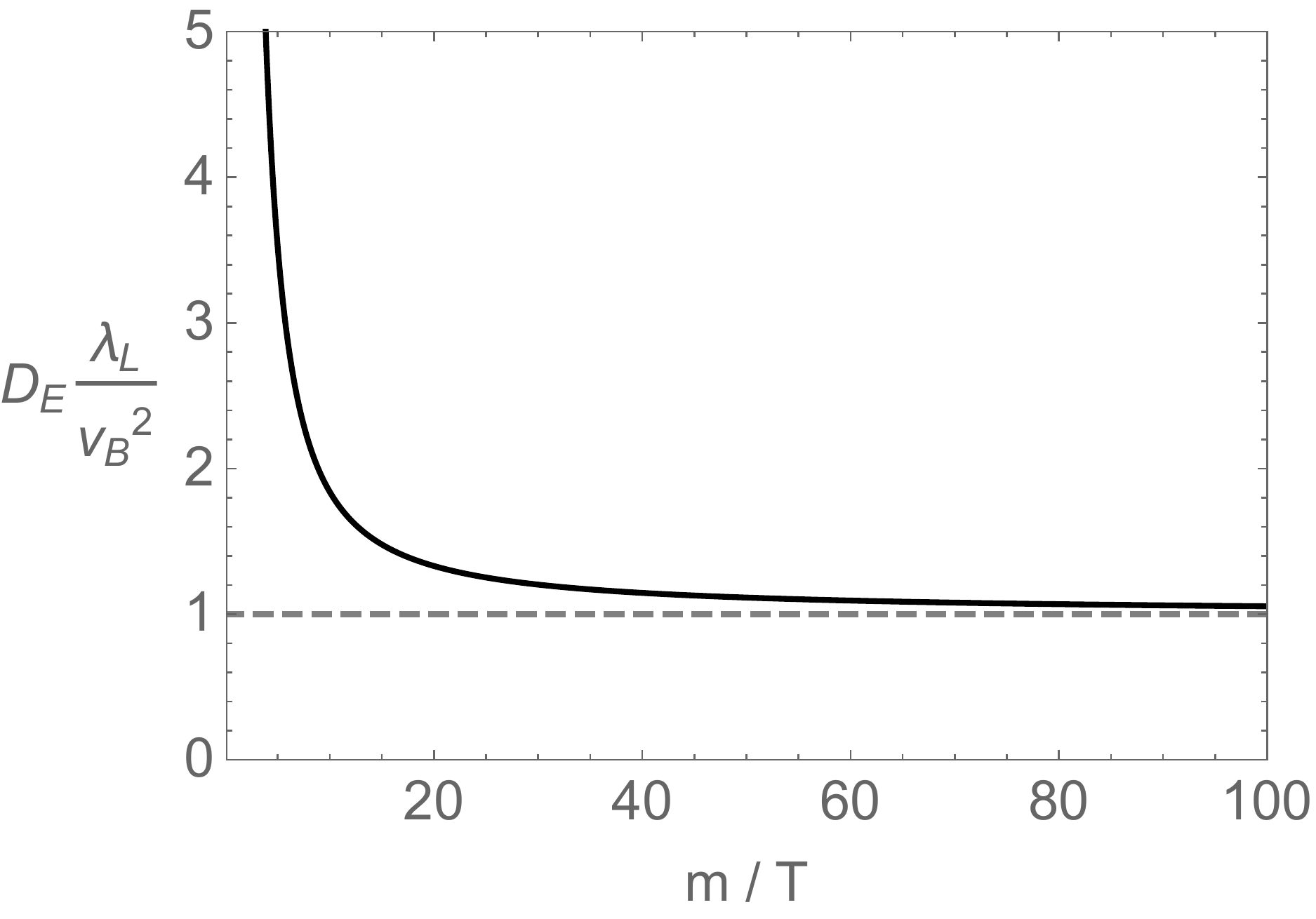} \label{}}
 \caption{The diffusion bounds of $D_{E}$ \eqref{nsgodxcew} with $N=1$. $\lambda_{L}=2\pi T$, the butterfly velocity $v_{B}$ is \eqref{VTFOMU22}. The dashed line denotes the lower bound \eqref{EDB22}.}\label{LBN1FIG}
\end{figure}

\paragraph{Pole-skipping and the energy diffusion with EXB:}
In the original study of pole-skipping~\cite{Blake:2018leo}, the following Lagrangian was used 
\begin{align}\label{PW122}
\begin{split}
\mathcal{L}_{M} = -\frac{1}{2} \sum_{i=1}^{2}(\partial\varphi_{i})^2 \,,\quad \varphi_i = m \, x^i \,,
\end{split}
\end{align}
which corresponds to the $N=1$ case in \eqref{genaxionmodel}.
For this model, one can find the energy diffusion constant is bounded from below \eqref{eq12first} as
\begin{align}\label{}
\begin{split}
D_{E} \,\frac{\lambda_{L}}{v_{B}^2} \geqslant 1\,,
\end{split}
\end{align} 
where the lower bound (an equality) is approached at low temperature ($m/T\gg1$). See Fig. \ref{LBN1FIG}.

With the quasi-normal mode computation, \cite{Blake:2018leo} showed that this lower bound can be also obtained from the pole-skipping as follows.
They found the pole-skipping point \eqref{PSP} passes though the energy diffusive dispersion relations 
\begin{align}\label{PW3123123}
\omega &= - i \, D_{E} \, k^2
\end{align} 
at the small temperature limit ($m/T\gg1$), in other words, we have
\begin{align}\label{EDB22}
\begin{split}
\omega_{*} = - i \, D_{E} \, k_{*}^2  \quad\rightarrow\quad D_{E} \frac{\lambda_{L}}{v_{B}^2} = 1\,.
\end{split}
\end{align}
This corresponds to the reproduction of the lower bound of energy diffusion in holography in terms of the poles-skipping phenomena.

%
%

The result \eqref{EDB22} also implies that, at $m/T\gg1$, we have: i) one can use the energy diffusion constant ($D_{E}$) to find the butterfly velocity ($v_{B}$); ii) the dispersion at the quadratic order \eqref{PW3123123} becomes the very good approximation to the exact location of the hydrodynamic poles even at ($\omega_{*}, k_{*}$). The second item, in particular, is not only the main reason for obtaining the bound \eqref{EDB22}, but also a non-trivial result.

In the next section, we will investigate how the above two results (i, ii) can be seen at the different symmetry breaking pattern: the SSB. 
For the parallel comparison with the SSB, we also review how one can obtain the hydrodynamic dispersion relation \eqref{PW3123123} for the EXB here. 

\paragraph{\textbf{Dispersion relation of the EXB}:}
The dispersion relation of the lowest modes in \eqref{PW2PW2PW2} is given by\footnote{This dispersion relation is supposed to be valid in the coherent regime, i.e., $\Gamma/T \ll 1$.}~\cite{Davison:2014lua}
\begin{align}
\omega = \pm \,k\, \sqrt{\frac{\partial P}{\partial \epsilon} - \frac{1}{4} \left( \frac{\Gamma}{k} + \frac{\eta}{\epsilon+P} \, k \right)^2 }  \,-\, \frac{i}{2}  \left(\Gamma + \frac{\eta}{\epsilon+P} \,k^2\right)  \,, \label{AXIONDISDIS}
\end{align}
where $\epsilon$ is the energy density, $P$ is the pressure, and $\eta$ is the shear viscosity.
$\Gamma$ sets a momentum dissipation rate as
\begin{align}
\Gamma  = \frac{s}{4\pi}\frac{m^2}{\epsilon+P} = \frac{m^2}{4\pi T}  \,. \label{GMMAAX}
\end{align}
The dispersion relation \eqref{AXIONDISDIS} can be rewritten in the small wave vector limit as follows.
\paragraph{\textbf{Without axion charge}:} it corresponds to the sound mode 
\begin{align}
(m=0): \quad \omega &= \pm \sqrt{\frac{\partial P}{\partial \epsilon}} k \,-\, i  \frac{\eta}{2(\epsilon+P)} \,k^2  \,, \label{axionm03weqwe} 
\end{align}
\paragraph{\textbf{With axion charge}:} it gives two diffusive modes 
\begin{align}\label{}
(m\neq0): \quad &\omega = -i  \frac{\partial P}{\partial \epsilon} \Gamma^{-1} \,k^2  \,\,\,\,=:\, -i \,D_{\text{Axion}}\, k^2 \,, \label{PW32} \\
(m\neq0): \quad &\omega = -i \, \Gamma   +  i \left(  \frac{\partial P}{\partial \epsilon}\Gamma^{-1}  -  \frac{\eta}{\epsilon+P}    \right) k^2         \,. \label{PW3PW3222} 
\end{align} 
Note that we define the diffusion constant in \eqref{PW32} as $D_\text{{Axion}}$ for later use and the second diffusive mode \eqref{PW3PW3222} is a pseudo-diffusive mode due to the finite frequency at zero wave vector.

In order to compare the dispersion relations of the lowest mode (\eqref{axionm03weqwe}-\eqref{PW3PW3222}) with the quasi-normal mode computation, we may need to specify several quantities ($T, \epsilon, P, s, \eta$) in holography.

The thermodynamic quantities ($T, \epsilon, P, s$)~\cite{Davison:2014lua} are given as 
\begin{align}
 T \,=\, \frac{r_{h}}{4\pi} \left( 3- \frac{m^2}{2 r_{h}^2} \right) \, \quad \epsilon \,=\, 2 \,r_{h}^3  - m^2 r_{h}   \,, \quad P \,= \, r_{h}^3 + \frac{m^2 \, r_{h}}{2} \,, \quad s \,=\, 4\pi r_{h}^2   \,,  \label{THMERRESULTS}
\end{align}
which is consistent with \eqref{THMERRESULTSGEN} at $N=1$, and the shear viscosity ($\eta$) can be computed numerically by following \cite{Hartnoll:2016tri}.
\begin{figure}[]
\centering
     \subfigure[$4\pi \eta/s$ vs $m/T$]
     {\includegraphics[width=6.5cm]{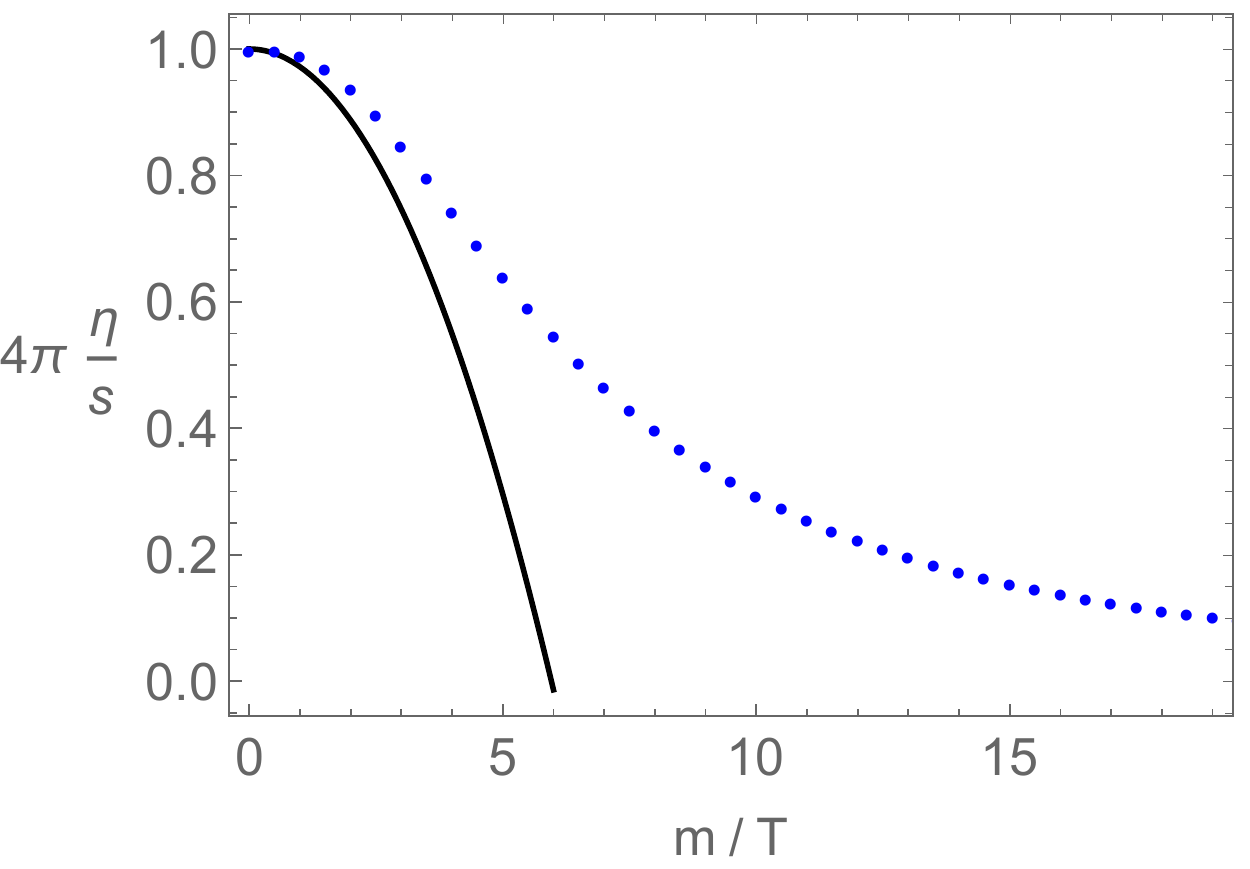} \label{shearfig}}
     \subfigure[Im($\omega$) at $m/T=1/2$]
     {\includegraphics[width=7.2cm]{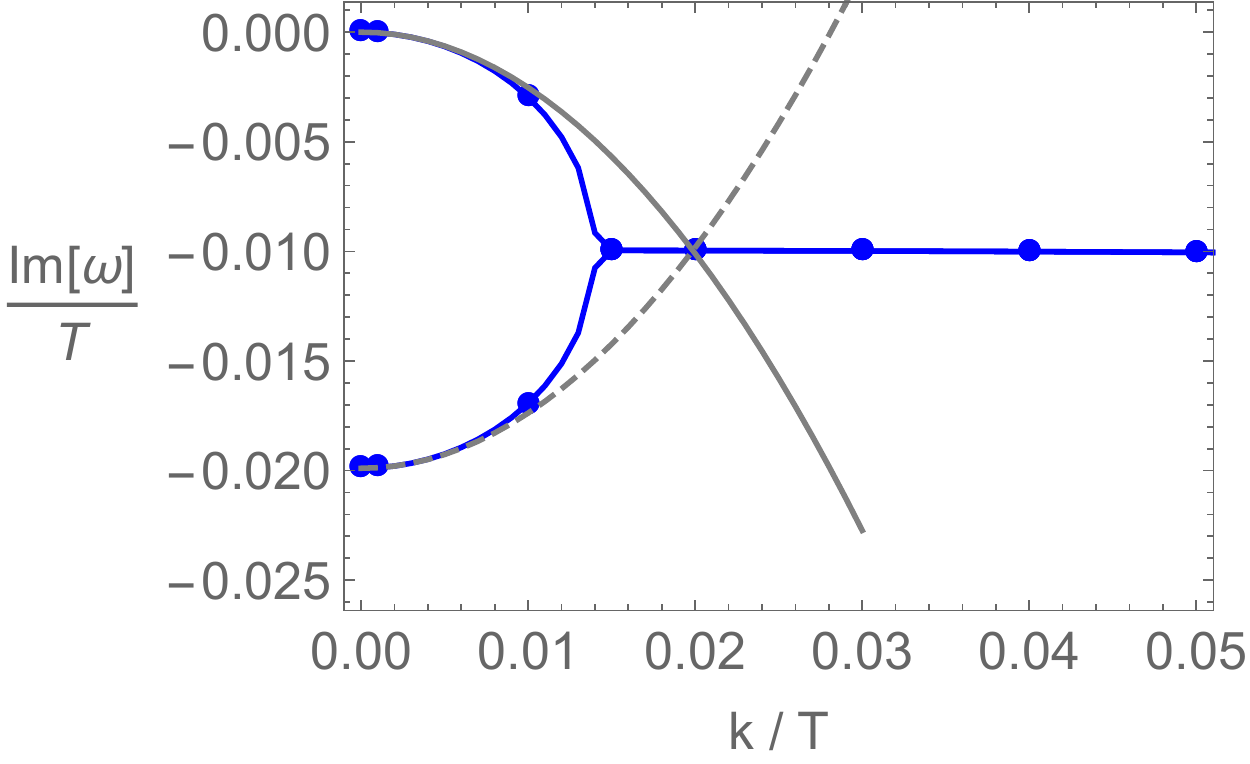} \label{AXIONQNM}}
 \caption{\textbf{Left:} The violation of KSS bound with axion charge. Blue dots are numerical results, and the black solid line is the analytic result of \eqref{SHEARANLY}. \textbf{Right:} Quasi-normal modes vs hydrodynamic predictions. Blue dots are numerically computed quasi-normal modes, the blue solid line is \eqref{AXIONDISDIS}, the gray solid line is \eqref{PW32}, and the gray dashed line is \eqref{PW3PW3222}. }\label{}
\end{figure}
We display the shear viscosity in Fig. \ref{shearfig}, and in particular, we display it in terms of the combination with entropy density. Blue dots are numerically computed data and the black solid line is the analytic expression valid in the small $m/T$ limit:
\begin{align}
4\pi \frac{\eta}{s} \,=\, 1 \,+\, \frac{\sqrt{3}}{16\pi} \left( 1 - \frac{3\sqrt{3} \log 3}{ \pi} \right) \frac{m^2}{T^2} \,. \label{SHEARANLY}
\end{align}
Including \eqref{SHEARANLY} with numerical data, Fig. \ref{shearfig} corresponds to the reproduction of results in \cite{Hartnoll:2016tri}. One can see the well-known breakdown of KSS bound by increasing the axion charge.

In Fig.\ref{AXIONQNM}, using the determinant method (see appendix \ref{appendixb}), we compute the quasi-normal modes and check that they match well to the hydrodynamic dispersion relations\footnote{We choose $m/T=1/2$ ($\Gamma/T \sim 0.02$) in order to match quasi normal modes with hydrodynamic dispersion relations \eqref{AXIONDISDIS} valid in $\Gamma/T\ll1$.}.
Blue dots are numerically computed quasi-normal modes and we have three dispersions: i) the blue solid line for \eqref{AXIONDISDIS}; ii) the gray solid line for \eqref{PW32}; iii) the gray dashed line for \eqref{PW3PW3222}. 
This figure is a reproduction of results in \cite{Davison:2014lua}.

\paragraph{Energy diffusion constant:}
Here let us make a comment on the diffusion constant in \eqref{PW32}, $D_{\text{Axion}}$. It can be replaced by the energy diffusion constant $D_{E} := \kappa/(T\partial s/\partial T)$\footnote{One can also see this replacement from equation (3.7) in ~\cite{Davison:2014lua}.}.
We can explicitly check this as follows.
First, we rewrite $D_{\text{Axion}}$ as
\begin{align}\label{DADE}
D_{\text{Axion}} := \frac{\partial{P}}{\partial{\epsilon}} \Gamma^{-1} \,=\, \frac{T}{m^2} \sqrt{4\pi^2 + \frac{3 m^2}{2 T^2}} \,,
\end{align}
where we use \eqref{GMMAAX}, \eqref{THMERRESULTS} in the second equality. 
Second, for the energy diffusion constant, we express $\kappa$ \eqref{nsgodxcew} in terms of ($T, m$)
%
\begin{align}
\kappa = \frac{4\pi^2 T}{9 m^2} \left( 4\pi T +\sqrt{6 m^2 + 16\pi^2 T^2} \right)^2 \,,
\end{align}
then we obtain
\begin{align}\label{DADE2}
D_{E} := \frac{\kappa}{T (\partial s/\partial T)} \,=\, \frac{T}{m^2} \sqrt{4\pi^2 + \frac{3 m^2}{2 T^2}} \,,
\end{align}
where we have used \eqref{THMERRESULTS} for ($s, T$). Compare \eqref{DADE} with \eqref{DADE2}, we see $D_{\text{Axion}}$ has the same form with the energy diffusion constant $D_{E}$\footnote{Note that \eqref{DADE2} is valid for any value of $m$, while \eqref{DADE} is obtained using a coherent result ($m/T \ll 1$). From the equivalence between \eqref{DADE} and \eqref{DADE2}, we may change $D_{\text{Axion}}$ in \eqref{PW32} as $D_{E}$.}.

\paragraph{Hydrodynamic mode of EXB:}
Based on the previous paragraph, we may summarize the hydrodynamic mode\footnote{Here the hydrodynamic mode is the one such that the frequency goes to zero at zero wave vector.} of EXB, \eqref{axionm03weqwe}-\eqref{PW32}, as
%
%
%
%
%
\begin{align}
(m=0): \quad \omega &= \pm \sqrt{\frac{\partial P}{\partial \epsilon}} k \,-\, i  \frac{\eta}{2(\epsilon+P)} \,k^2  \,,\label{m0} \\
(m\neq0): \quad \omega &  \,=\, - i \, D_{E} \, k^2 \,,  \label{NANANANA}
\end{align} 
where $m$ could be any non-zero value\footnote{In the incoherent regime ($m/T\gg1$), the quasi-normal modes follow the energy diffusion modes~\cite{Davison:2014lua}.}.
This is how the dispersion relation \eqref{PW3123123} is described in a previous work~\cite{Blake:2018leo} and they showed the pole-skipping point \eqref{PSP} passes through \eqref{NANANANA} at the small temperature limit ($m/T\gg1$).

\subsection{Spontaneous symmetry breaking}
The dispersion relation of the lowest modes in \eqref{PW2PW2PW2} depends on the symmetry breaking patterns. In other words, \eqref{NANANANA} cannot be true if the symmetry breaking pattern changes.

When the symmetry is broken spontaneously, we obtain the longitudinal sound modes (phononic vibrational modes)~\cite{Ammon:2019apj,Andrade:2017cnc,Baggioli:2020ljz}
\begin{align}
\omega = \pm v_{L} \,k \,-\, i \, D_{L} \, k^2  \,,  \label{SSSSSSSMO}
\end{align} 
where $v_{L}$ denotes the sound speed and $D_{L}$ is the diffusion constant. This mode is related to the properties of materials such that the coefficients ($v_{L}, D_{L}$) are functions of viscous coefficients (shear and bulk elastic moduli, shear viscosity, etc)\footnote{The sound mode \eqref{SSSSSSSMO} becomes \eqref{m0} in the absence of the translational symmetry breaking ($m=0$).}.

In addition to this sound mode, there is an additional diffusive mode (the crystal diffusion mode)~\cite{Ammon:2019apj,Andrade:2017cnc,Baggioli:2020ljz,Baggioli:2020nay,Armas:2019sbe,Ammon:2020xyv} for the SSB with the dispersion relation:
\begin{align}
\omega = -i D_{\phi} \, k^2 \,,  \label{ADDDDD}
\end{align} 
where the $D_{\phi}$ is called the crystal diffusion constant related to the Goldstone mode, which is a function of several quantities (e.g., the Goldstone dissipative parameter, bulk modulus, shear viscosity, and crystal pressure, etc)\footnote{The physical nature of this diffusive mode is still controversial~\cite{Baggioli:2020ljz}. We refer the reader to \cite{Baggioli:2020nay} for the discussion on a possible interpretation in terms of quasicrystal physics.}.
This $D_{\phi}$ is the different quantity from the energy diffusion constant $D_{E}$. For instance, $D_{E}$ is divergent in $m/T\ll1$ limit while $D_{\phi}$ is still finite.

\begin{figure}[]
\centering
     \subfigure[$v_{L}$]
     {\includegraphics[width=4.63cm]{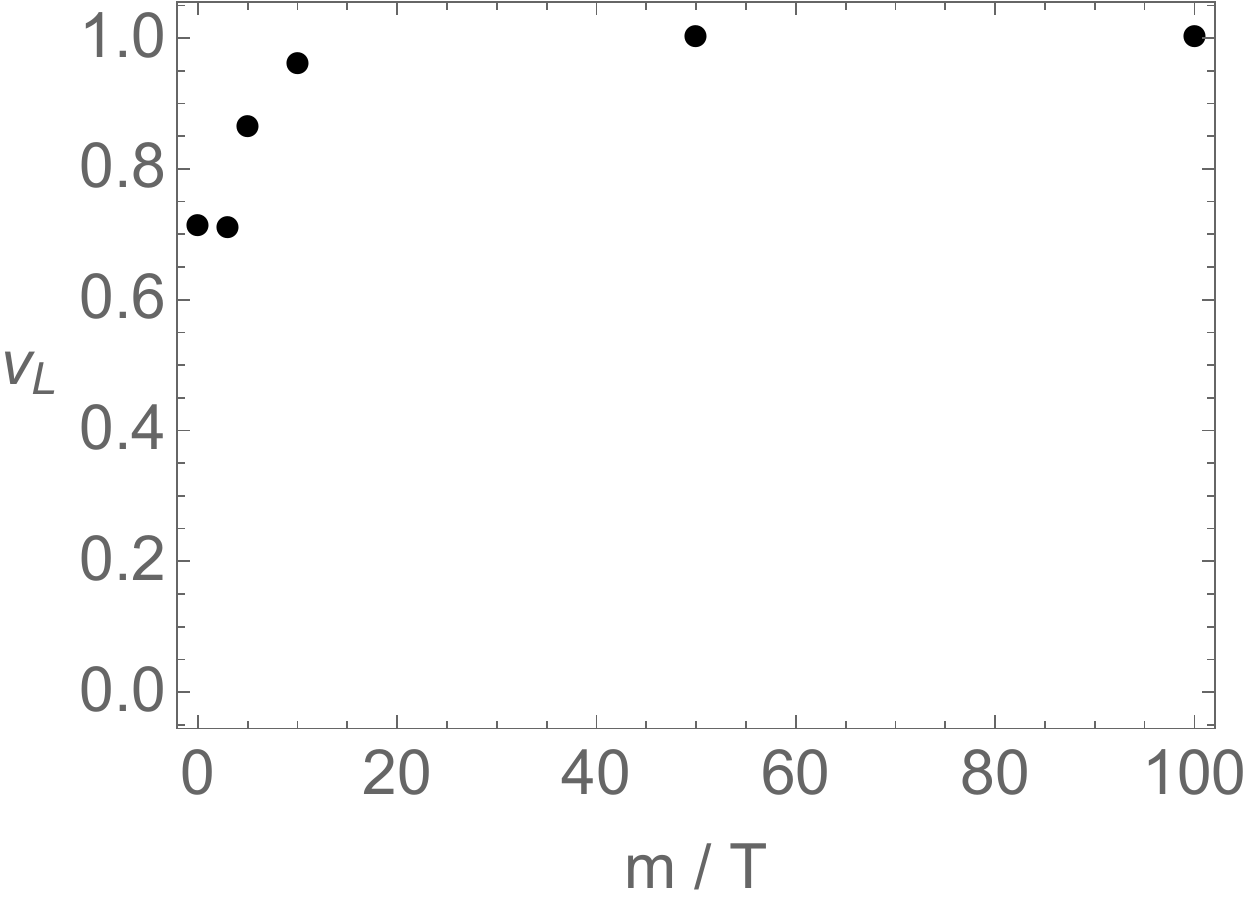} \label{}}
     \subfigure[$D_{L}$]
     {\includegraphics[width=4.83cm]{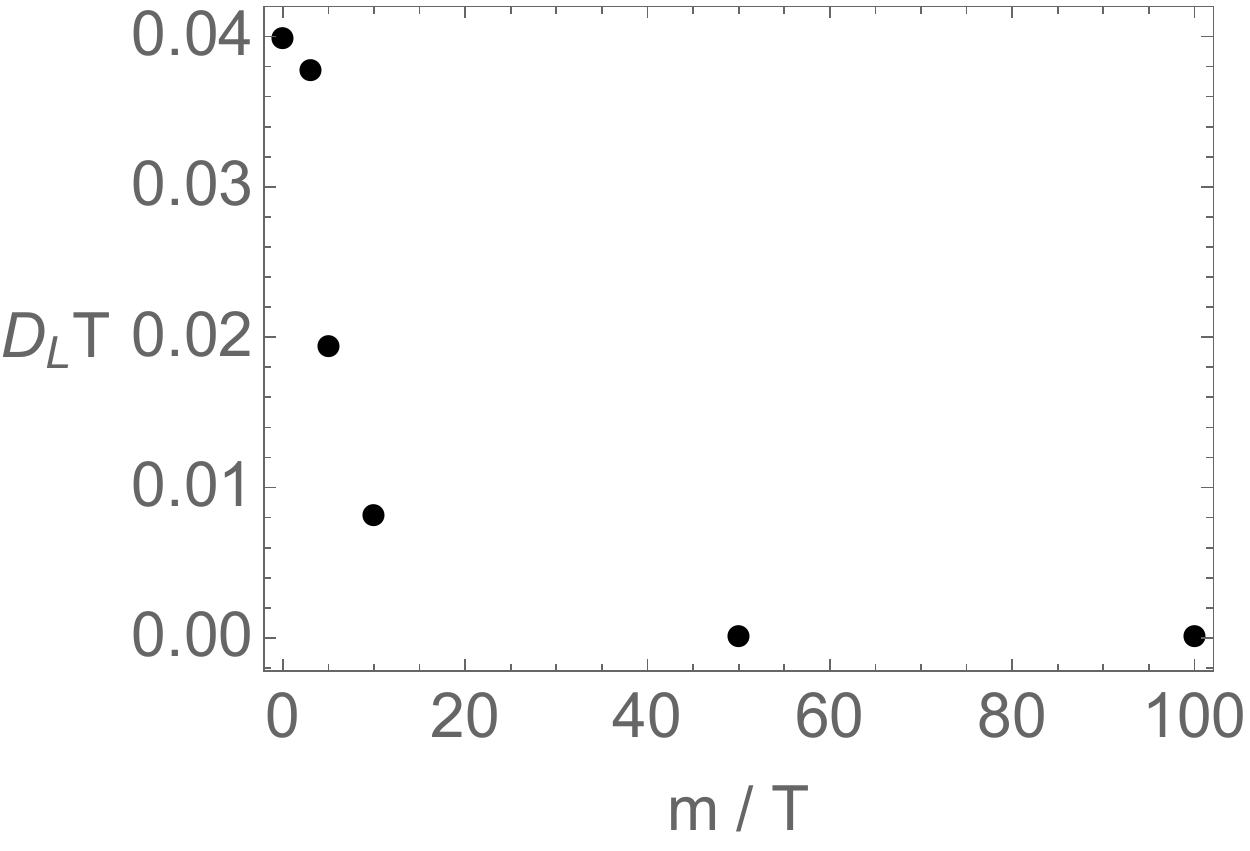} \label{}}
     \subfigure[$D_{\phi}$]
     {\includegraphics[width=4.83cm]{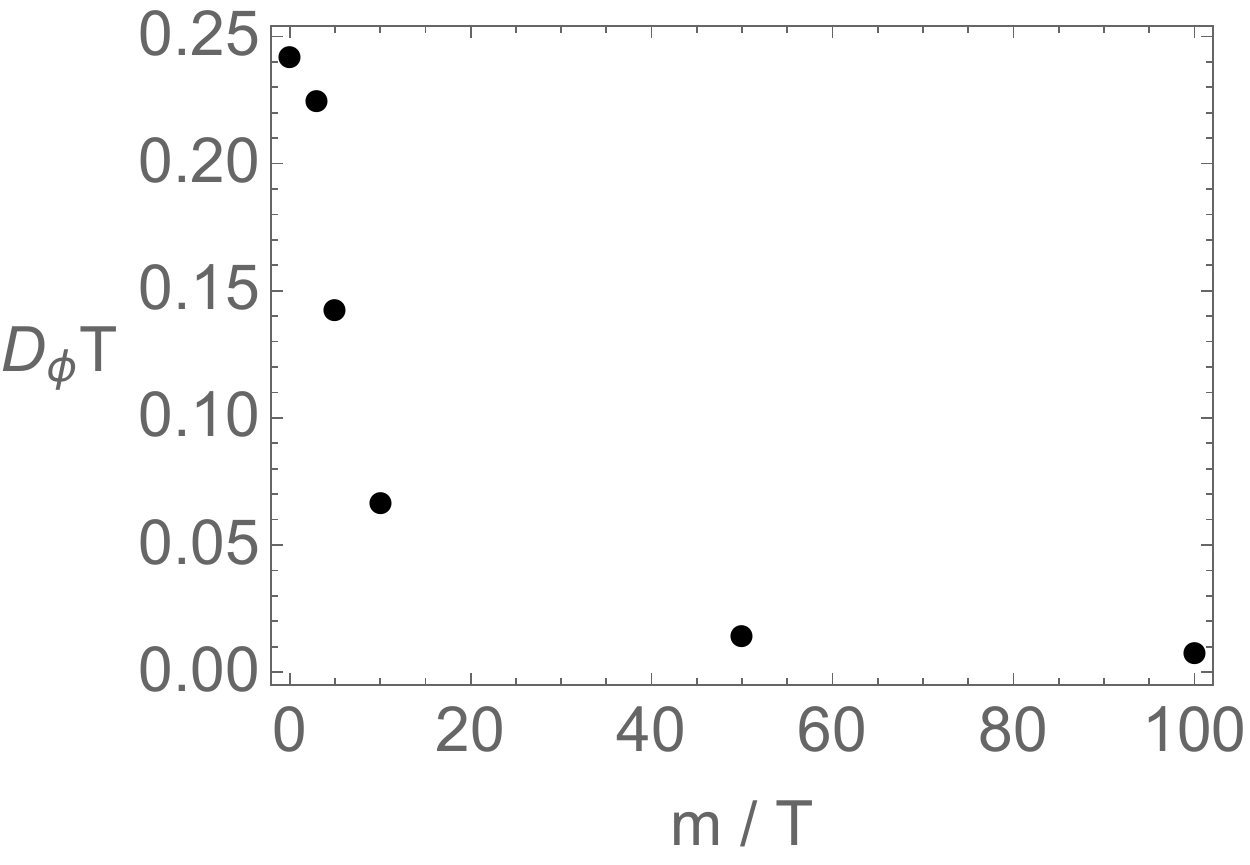} \label{}}
 \caption{Coefficients of dispersion relations of the SSB, ($v_{L}, D_{L}, D_{\phi}$), with various $m/T$.}\label{SSBREPRO}
\end{figure}

We choose $N=3$ in \eqref{genaxionmodel} as a toy model for the SSB case. Using the determinant method, we compute the quasi-normal mode spectrum of this model and then we numerically determine the corresponding coefficients in the hydrodynamic regime\footnote{We read them off by the fitting in the small wave number regime ($k/T<1$).}: ($v_{L}, D_{L}, D_{\phi}$) in \eqref{SSSSSSSMO}, \eqref{ADDDDD}.
In Fig. \ref{SSBREPRO}, we make a plot of it, which is consistent with \cite{Ammon:2019apj}. 
For the comparison between numerics and analytic results, we refer the reader to \cite{Ammon:2019apj,Baggioli:2020ljz}.

From Fig. \ref{SSBREPRO}, one can observe how the coefficients of the SSB, ($v_{L}, D_{L}, D_{\phi}$), change with respect to the temperature.
As the temperature is lowered ($m/T\gg1$), $v_{L}$ reaches a maximum value\footnote{Note that $v_{L}$ begins from the conformal value $v_{L} = v_{c} = 1/\sqrt{2}$ at $m=0$. For an arbitrary spacetime dimension $d$, the conformal theory has the speed of sound $v_{c}^2=\frac{c^2}{d-1}$ where $c$ is the speed of light. Note also the maximum value depends on the value of $N$~\cite{Baggioli:2020ljz}.} and the two diffusion constants ($D_{L}\,T,\, D_{\phi}\,T$) vanish.

\paragraph{The diffusion bound with the crystal diffusion.} 
Following the parallel procedure of the EXB case in Fig. \ref{LBN1FIG}, one can also study the diffusion bounds with these two diffusion constants ($D_{L} \,, D_{\phi}$) of the SSB case as
\begin{align}\label{}
\begin{split}
D_{L} \frac{\lambda_{L}}{v_{B}^2} \,, \quad D_{\phi} \frac{\lambda_{L}}{v_{B}^2} \,, 
\end{split}
\end{align} 
and if one can find the diffusion bound with them, we may check if those bounds also can be captured from the pole-skipping phenomena as in \eqref{EDB22}.

\begin{figure}[]
\centering
     \subfigure[Checking the diffusion bound for $D_{L}$]
     {\includegraphics[width=7.2cm]{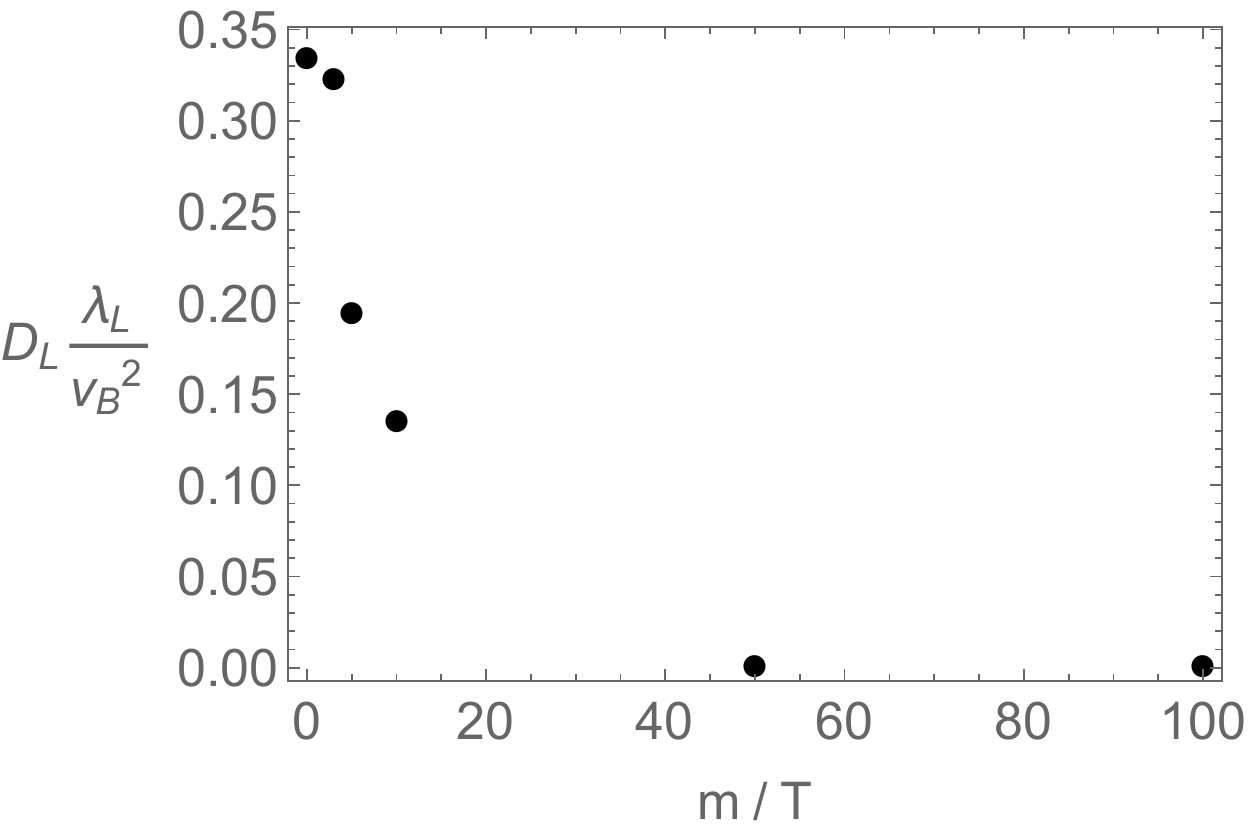} \label{FIGEDDDa}}
     \subfigure[Checking the diffusion bound for $D_{\phi}$]
     {\includegraphics[width=7.2cm]{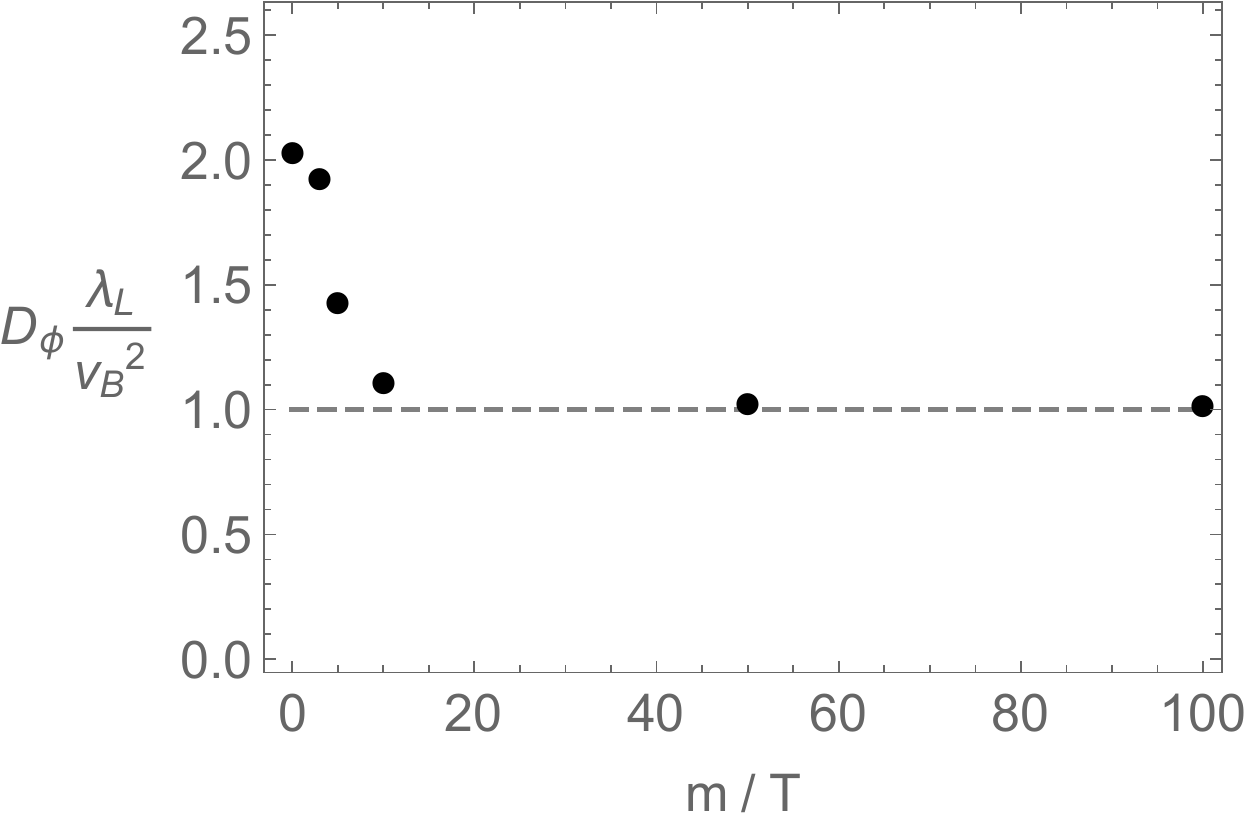} \label{FIGEDDDb}}
 \caption{The diffusion bounds of the SSB case for $D_{L}$ and $D_{\phi}$. $\lambda_{L}=2\pi T$, the butterfly velocity $v_{B}$ is \eqref{VTFOMU22}. The dashed line denotes the lower bound.}\label{FIGEDDD}
\end{figure}

In~\cite{Baggioli:2020ljz}, it turned out that $D_{L}$ cannot make the diffusion bound (see Fig. \ref{FIGEDDDa}), it is vanishing at low temperature. 
On the other hand, the crystal diffusion constant, $D_{\phi}$, has the diffusion lower bound  (see Fig. \ref{FIGEDDDb}) as
\begin{align}\label{DPHIBOUND}
\begin{split}
D_{\phi} \frac{\lambda_{L}}{v_{B}^2} \,\geqslant\, 1\,,
\end{split}
\end{align} 
where the equality (the lower bound) is approached at low temperature ($m/T\gg1$).
%

\paragraph{Pole-skipping and the crystal diffusion.} 
In what follows, we study the quasi-normal modes of the SSB with the pole-skipping, showing that the lower bound with crystal diffusion can be captured by the pole-skipping point.

\begin{figure}[]
\centering
     \subfigure[$m/T=5$]
     {\includegraphics[width=4.83cm]{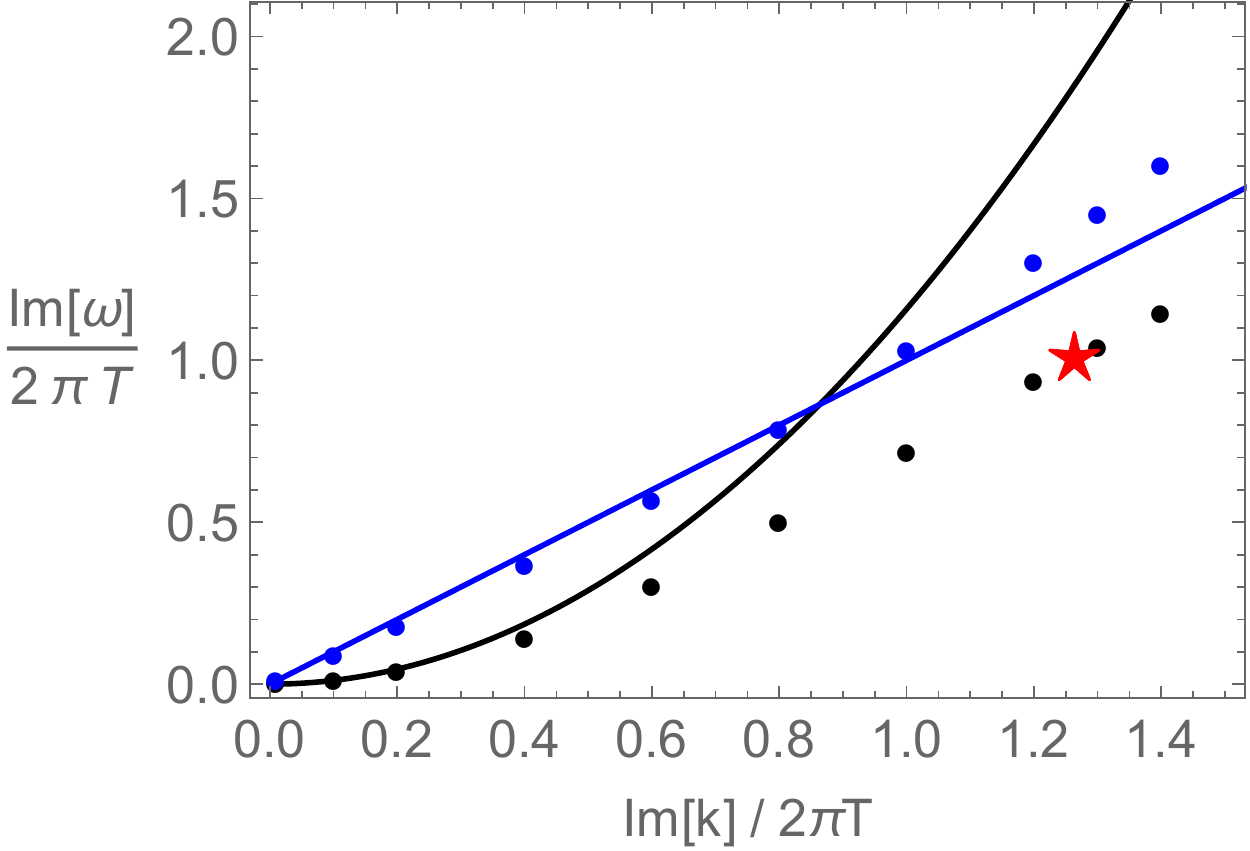} \label{}}
     \subfigure[$m/T=10$]
     {\includegraphics[width=4.83cm]{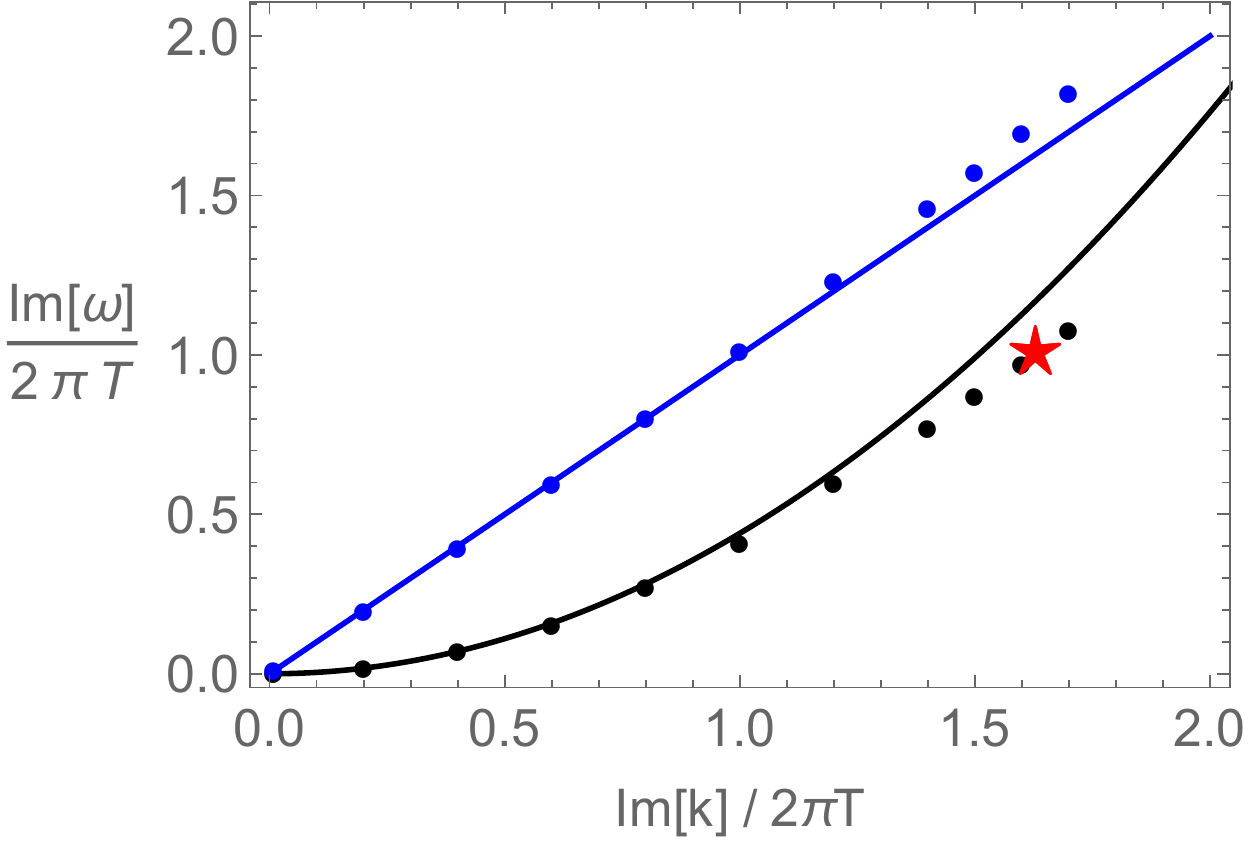} \label{}}
     \subfigure[$m/T=100$]
     {\includegraphics[width=4.63cm]{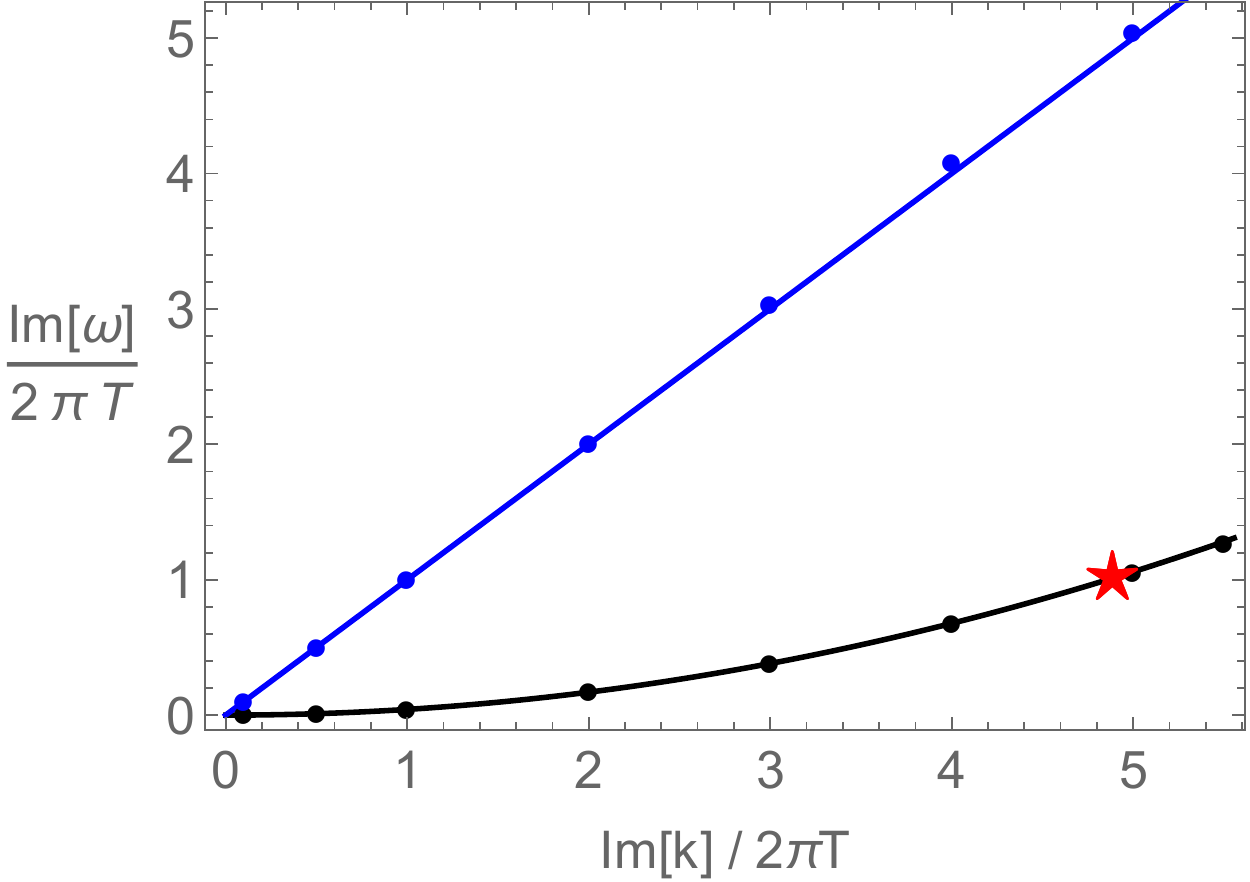} \label{}}
 \caption{Pole-skipping and quasi-normal modes in the SSB case with varying $m/T$. In all figures, dots are numerically computed quasi-normal modes, the red star is the pole-skipping point \eqref{PSP}, the black solid line is the crystal diffusion mode in \eqref{ADDDDD}, and the blue solid line is the longitudinal sound mode \eqref{SSSSSSSMO}.}\label{PSSSBFIG}
\end{figure}

In Fig. \ref{PSSSBFIG}, we display four things at different $m/T$: i) numerically computed quasi-normal modes (dots); ii) the pole-skipping point (red star) \eqref{PSP}; iii) the longitudinal sound mode (blue solid line) \eqref{SSSSSSSMO}; iv) the crystal diffusion mode (black solid line) \eqref{ADDDDD}.

From this figure, we can see that quasi-normal modes are well matched with hydrodynamic dispersion relations of the SSB (\eqref{SSSSSSSMO}, \eqref{ADDDDD}) in the small ($\omega, k$) regime at given temperature.
Moreover, as the temperature is lowered from Fig. \ref{PSSSBFIG}(a)  to Fig. \ref{PSSSBFIG}(c), one can notice that hydrodynamic dispersions become a good approximation for quasi-normal modes even in the large ($\omega, k$) regime.

This low-temperature feature also shows that the pole-skipping point \eqref{PSP} (red star) passes through the crystal diffusion mode (black line), see Fig. \ref{PSSSBFIG}(c), in other words,  
\begin{align}\label{}
\begin{split}
\omega_{*} = - i \, D_{\phi} \, k_{*}^2  \quad\rightarrow\quad D_{\phi} \frac{\lambda_{L}}{v_{B}^2} = 1\,,
\end{split}
\end{align}
which is analogous to \eqref{EDB22}, connecting $D_{\phi}$ with the butterfly velocity $v_{B}$.

Thus our work shows that, in addition to the EXB case (energy diffusion bound), the pole-skipping also can capture the diffusion bound (crystal diffusion bound) for the SSB case in \eqref{DPHIBOUND}\footnote{Our work also supports the hypothesis given in \cite{Wu:2021mkk}: $D\ge\frac{v_{skip}^2}{\omega_{skip}}$ where $v_{skip}:=v_{B}, \,\omega_{skip}:=\lambda_{L}$.}.

%
\section{Pole-skipping and magnetic fields}
In this section, we take another way (without axion fields) to study the EXB of translational invariance in holography: we break the translational invariance using magnetic fields.
In particular, with the quasi-normal mode computations, we will show explicitly that the pole-skipping point passes through the magneto-hydrodynamic dispersion with strong magnetic fields (or low temperature limit). 

From the perspective of hydrodynamics (effective theories), this result might be expected because when momentum is relaxed, there would be a parameter for the momentum dissipation rate $\Gamma$. Then the system might be governed by the energy diffusion in the strong momentum relaxation regime ($\Gamma/T\gg1$), although the specific form of $\Gamma$ may depend on different systems (e.g., the axion model \eqref{PW122}, the magnetic field model \eqref{ACTION}).

However, matching quasi-normal modes from the gravity calculation with the hydrodynamic prediction (without holography) might be a non-trivial procedure. In particular, as we will show that, in order to correctly match the quasi-normal modes with magnetic fields, we need to consider the shear viscosity in the magneto-hydrodynamics, i,e., our work will not only explicitly show the expectation above, but also give a non-trivial consistency check for the viscous magneto-hydrodynamics in holography~\cite{Buchbinder:2008dc,Buchbinder:2009aa,Hansen:2009xe,Buchbinder:2009mk,Hartnoll:2007ip,Hansen:2008tq,Hartnoll:2007ih}\footnote{{For the recent development of magneto-hydrodynamics, see also \cite{Amoretti:2021fch,Amoretti:2020mkp,Amoretti:2019buu} and references therein.}} .

%
\subsection{Holographic model of EXB with magnetic fields}

In this section we break the translational symmetry explicity in holography using the following magnetically charged model 
\begin{align}\label{ACTION}
\begin{split}
\mathcal{L}_{M} = -\frac{1}{4} F^2 \,, \quad    A = -\frac{H}{2} y \,\dd x \,+\, \frac{H}{2} x \, \dd y \,,
\end{split}
\end{align}
where $F=\dd A$ is the field strength of the gauge field $A$, which corresponds to \eqref{GENCLASS} with
\begin{align}\label{}
\begin{split}
\phi = 0 \,,\quad  V(\phi) = 0  \,,\quad  Z(\phi) = 1 \,,\quad Y(\phi) = 0 \,,\quad W(X) = 0 \,,
\end{split}
\end{align}
and
\begin{align}\label{}
\begin{split}
A_{v} = 0 \,,\quad H \neq 0\,.
\end{split}
\end{align}

Then, this action yields the equation of motion
\begin{align}
\nabla_{\mu}F^{\mu\nu}  = 0 \,, \quad
R_{\mu\nu} - \frac{1}{2} g_{\mu\nu} \left[R + 6 - \frac{1}{4} F^2 \right] = \frac{1}{2} F_{\mu\d}F_{\nu}{^\d} \,,
\end{align}
and the background solution \eqref{METANS} is
\begin{equation}\label{bgc}
\begin{split}
 D(r)\,= \frac{1}{B(r)} = r^2 - \frac{m_{0}}{r} \,+ \, \frac{H^2}{4\,r^2} \,, \quad C_{1}(r) = C_{2}(r) =  r^2 \,,
\end{split}
\end{equation}
where $m_{0}$ is determined by the condition $D(r_{h})=0$:
\begin{align}\label{m0m0m0}
m_{0} = r_{h}^3\left( 1 +  \frac{H^2}{4\, r_{h}^4} \right)\,, \qquad r_{h} = \text{ the radius of horizon,}
\end{align}
and the butterfly velocity is the same as \eqref{VTFOMU22}.

Including the Hawking temperature \eqref{HT}, thermodynamic quantities~\cite{Kim:2015wba} read
\begin{align}\label{HAWKINGT}
 T \,=\, \frac{1}{4\pi} \left( 3\,r_{h} \,-\, \frac{H^2}{4\,r_{h}^3}  \right) \, \quad \epsilon \,=\, 2 \,r_{h}^3  + \frac{H^2}{2\,r_{h}}   \,, \quad P \,= \, r_{h}^3 - \frac{3 H^2}{4 \, r_{h}} \,, \quad s \,=\, 4\pi r_{h}^2   \,.
\end{align}
Note that these quantities satisfy the following Smarr-like relation
\begin{align}\label{}
 \epsilon + P = s T \,.
\end{align}

\subsection{Fluctuations for quasi-normal modes}

In order to study the holographic dual of hydrodynamic modes in the action \eqref{ACTION}, we need to consider fluctuations on the black hole background as
\begin{align}\label{}
\begin{split}
g_{\mu\nu} \,\rightarrow\, g_{\mu\nu} + \delta g_{\mu\nu} \,, \quad A_{\mu} \,\rightarrow\, A_{\mu} + \delta A_{\mu} \,,
\end{split}
\end{align} 
At the linearized fluctuation level of the Einstein equations and the Maxwell equation, there are two sets of decoupled fluctuations:
\begin{align}\label{CHNNELCLASSIFICATION}
\begin{split}
&\text{(Sound channel):} \quad\,\,\,\,  \{\delta g_{tt}, \,\delta g_{tx}, \,\delta g_{xx}, \, \delta g_{yy},\, \delta A_{y}\} \,, \\
&\text{(Shear channel):} \qquad  \{\delta g_{ty}, \, \delta g_{xy}, \,\delta A_{t}, \, \delta A_{x}\} \,. \\
\end{split}
\end{align} 
In the field theory language, the first (second) set corresponds to the sound (shear) channel\footnote{Note that this decoupling is related to the $\mathbb{Z}_{2}$ parity symmetry along the $y$-axis~\cite{Buchbinder:2008dc,Buchbinder:2009aa}.}. 

Therefore, one can see that the sound channel in \eqref{CHNNELCLASSIFICATION} would be the main channel we focus in this paper, i.e., the sound channel is the gravitational sound mode \eqref{GSM} coupled with the matter fluctuation \eqref{MFF} $\delta \Phi = \delta A_{y}$ as :
\begin{align}\label{FLUCOURSETUP}
\begin{split}
\delta g_{tt} &= h_{tt}(r) \,e^{-i \, \omega \, t + i \, k \, x} \,,\, \quad  \delta g_{tx} = h_{tx}(r) \,e^{-i \, \omega \, t + i \, k \, x} \,,  \quad
\delta g_{xx} = h_{xx}(r) \,e^{-i \, \omega \, t + i \, k \, x} \,, \\
\delta g_{yy} &= h_{yy}(r) \,e^{-i \, \omega \, t + i \, k \, x} \,, \quad\delta A_{y} = a_{y}(r) \,e^{-i \, \omega \, t + i \, k \, x} \,.
\end{split}
\end{align} 
%

%
\subsection{Quasi-normal mode vs Magneto-hydrodynamics}
When the external magnetic field ($H$) is considered, we may expect that the quasi-normal modes from holography can be compared with the one from magneto-hydrodynamics. 
One can see the brief review of the magneto-hydrodynamics in appendix \ref{appendixa}, including both the sound channel and shear channel. Let us collect the main results of magneto-hydrodynamic of the sound channel which is the main channel considered in this paper as follows.

\paragraph{Without the magnetic field:}
in the absence of $H$, the magneto-hydrodynamic mode is the propagating damped sound mode
\begin{align}\label{SS1}
\begin{split}
(H=0): \quad \omega = \pm \sqrt{\frac{\partial P}{\partial \epsilon}} k \,-\, i  \frac{\eta}{2(\epsilon+P)} \,k^2 \,,
\end{split}
\end{align}
where $\epsilon$ is the energy density, $P$ is the pressure, and $\eta$ is a shear viscosity.  Note that this is the same as \eqref{axionm03weqwe}.

\paragraph{With the magnetic field:}
on the other hand, at finite $H$, the magneto-hydrodynamic modes give two diffusive modes\footnote{We have obtained the quadratic term in \eqref{SD1}, which did not appear in \cite{Buchbinder:2008dc,Buchbinder:2009aa}.} as,
\begin{align}\label{}
(H\neq0): \quad &\omega = -i  \frac{\partial P}{\partial \epsilon} \Gamma^{-1} \,k^2  \,=:\, -i \,D_{MHD}\, k^2 \,, \label{SD2} \\
(H\neq0): \quad &\omega = -i \Gamma   +  i \left(  \frac{\partial P}{\partial \epsilon}\Gamma^{-1}  -  \frac{\eta}{\epsilon+P}    \right) k^2         \,, \label{SD1} 
\end{align} 
where the momentum relaxation rate $\Gamma$\footnote{This $\Gamma$ also can be understood from the hydrodynamic cyclotron mode. See appendix \ref{appendixc}.} is 
\begin{align}\label{GMMAAXMFC}
\begin{split}
\Gamma = \sigma_{Q} \frac{H^2}{\epsilon + P}  \,.
\end{split}
\end{align} 
with a first-order transport coefficient $\sigma_{Q}$. 
{From the perspective of a comparison with axion model results \eqref{PW32}-\eqref{PW3PW3222}, magneto-hydrodynamic modes \eqref{SD2}-\eqref{SD1}\footnote{{Note that magneto-hydrodynamic modes at finite $H$ may depend on the behavior of $H$. For instance, one may study the sound waves if $H$ scales with a wave vector $k$ as $H\sim k$~\cite{Buchbinder:2008dc,Buchbinder:2009aa}. We thank Navid Abbasi for pointing this out.}} may be only valid at $\Gamma/T\ll1$ limit  ($H/T^2\ll1$)\footnote{{As did in a comparison between \eqref{DADE} and \eqref{DADE2}, we will show $D_{MHD}$ defined in a coherent regime can be replaced by an energy diffusion constant $D_{E}$.}}, which will be verified from the quasi-normal mode computations in short.}
One can compare $\Gamma$ in \eqref{GMMAAXMFC} with the axion model result \eqref{GMMAAX}.



\paragraph{Holographic computation:}
In order to compare the dispersion relations of the magneto-hydrodynamics, \eqref{SS1}-\eqref{SD1}, with the quasi-normal mode spectrum, we need to identify several quantities $\left(\eta, \,\epsilon + P, \, \frac{\partial{P}}{\partial{\epsilon}} \,, \sigma_{Q} \right)$ in holography.
The shear viscosity ($\eta$) can be given
\begin{align}\label{EQSET1}
\begin{split}
\eta \,=\, \frac{1}{4\pi} s \,=\,  r_{h}^2 \,,
\end{split}
\end{align} 
where the first equality implies that the KSS bound~\cite{Kovtun:2004de} holds in the presence of the external magnetic field~\cite{Buchbinder:2009aa} and \eqref{HAWKINGT} is used in the second equality.
The energy density ($\epsilon$) and the pressure  ($P$) are given in \eqref{HAWKINGT}, and then we have
\begin{align}\label{SMARR}
\begin{split}
\epsilon + P \,=\, 3 \,r_{h}^3 - \frac{H^2}{4\, r_{h}}  \,, \qquad  \frac{\partial{P}}{\partial{\epsilon}} = \frac{1}{2} - \frac{2\,H^2}{H^2 - 12 \,r_{h}^4} \,.
\end{split}
\end{align} 
The first-order transport coefficient $\sigma_{Q}$ is given by the fluid/gravity correspondence \cite{Blake:2015hxa}\footnote{{From the recent development of magneto-transport \cite{Amoretti:2021fch,Amoretti:2020mkp,Amoretti:2019buu}, \eqref{EQSET3} may have the correction if the magnetic field is no longer taken to be of order one in derivatives. However, we may use \eqref{EQSET3} for our purpose ($\Gamma/T\ll1$). We thank Daniel K. Brattan, Andrea Amoretti for pointing this out.}}
\begin{align}\label{EQSET3}
\begin{split}
\sigma_{Q} = \left(\frac{s T}{\epsilon + P}\right)^2 \,.
\end{split}
\end{align} 
%

In Fig. \ref{PUREEXBFIG}, using the determinant method (see appendix \ref{appendixb}), we display the quasi-normal modes together with the magneto-hydrodynamic dispersion relations.
\begin{figure}[]
\centering
     \subfigure[Re($\omega$) at $H/T^2=0$]
     {\includegraphics[width=4.83cm]{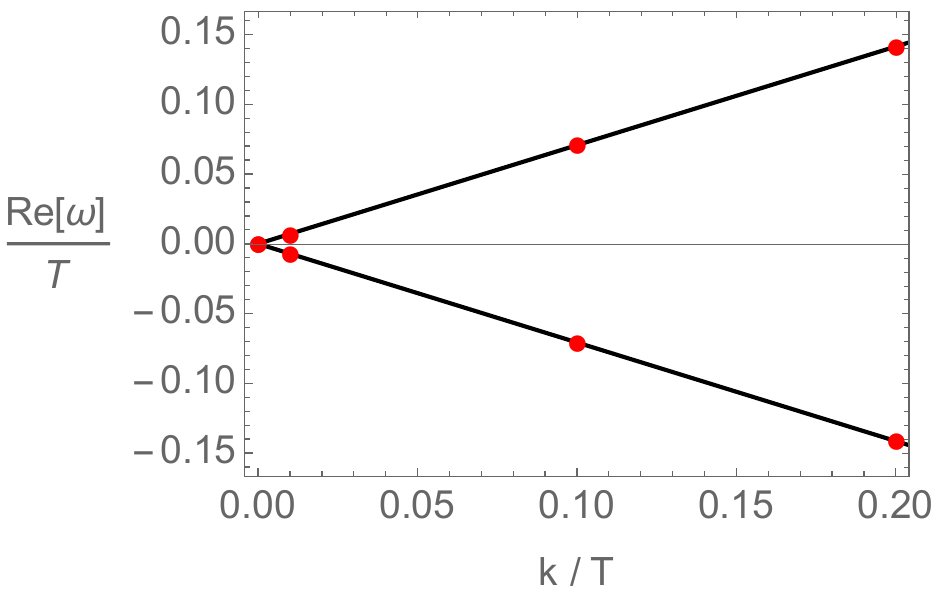} \label{}}
     \subfigure[Re($\omega$) at $H/T^2=3$]
     {\includegraphics[width=4.83cm]{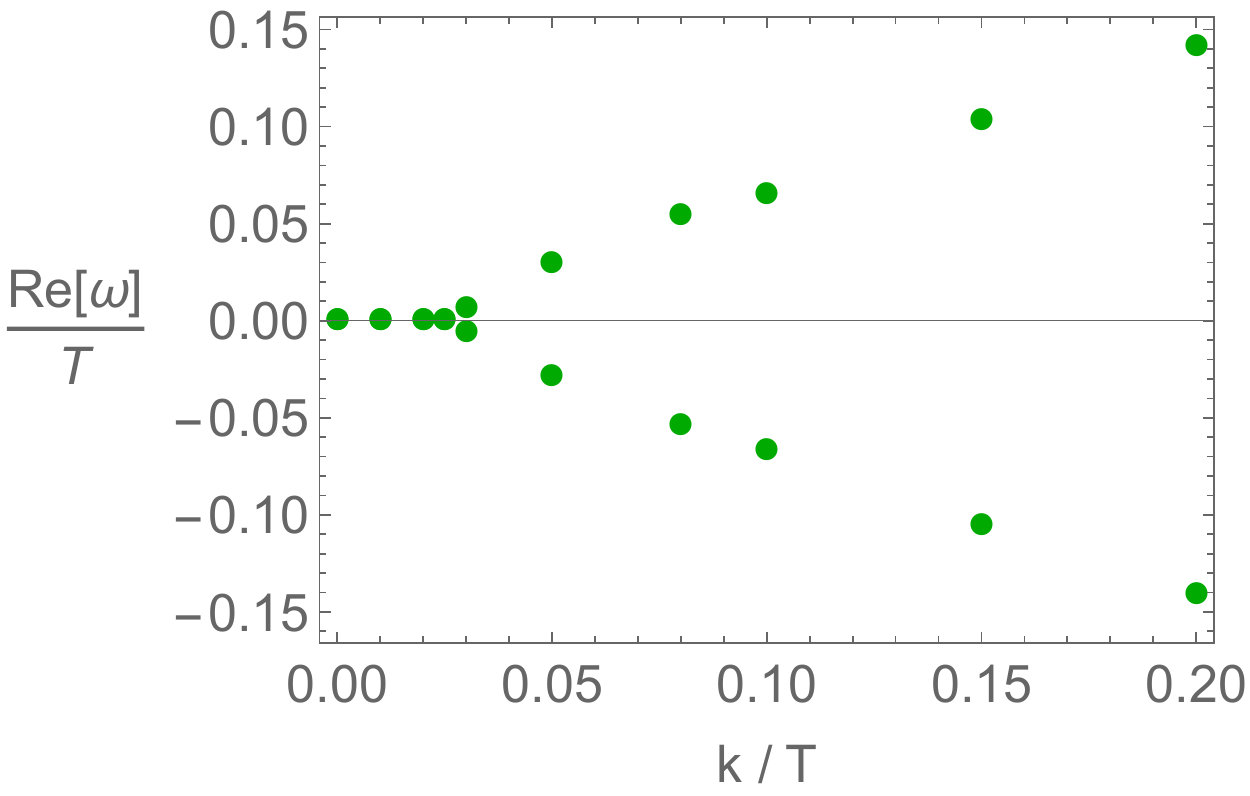} \label{}}
     \subfigure[Re($\omega$) at $H/T^2=5$]
     {\includegraphics[width=4.83cm]{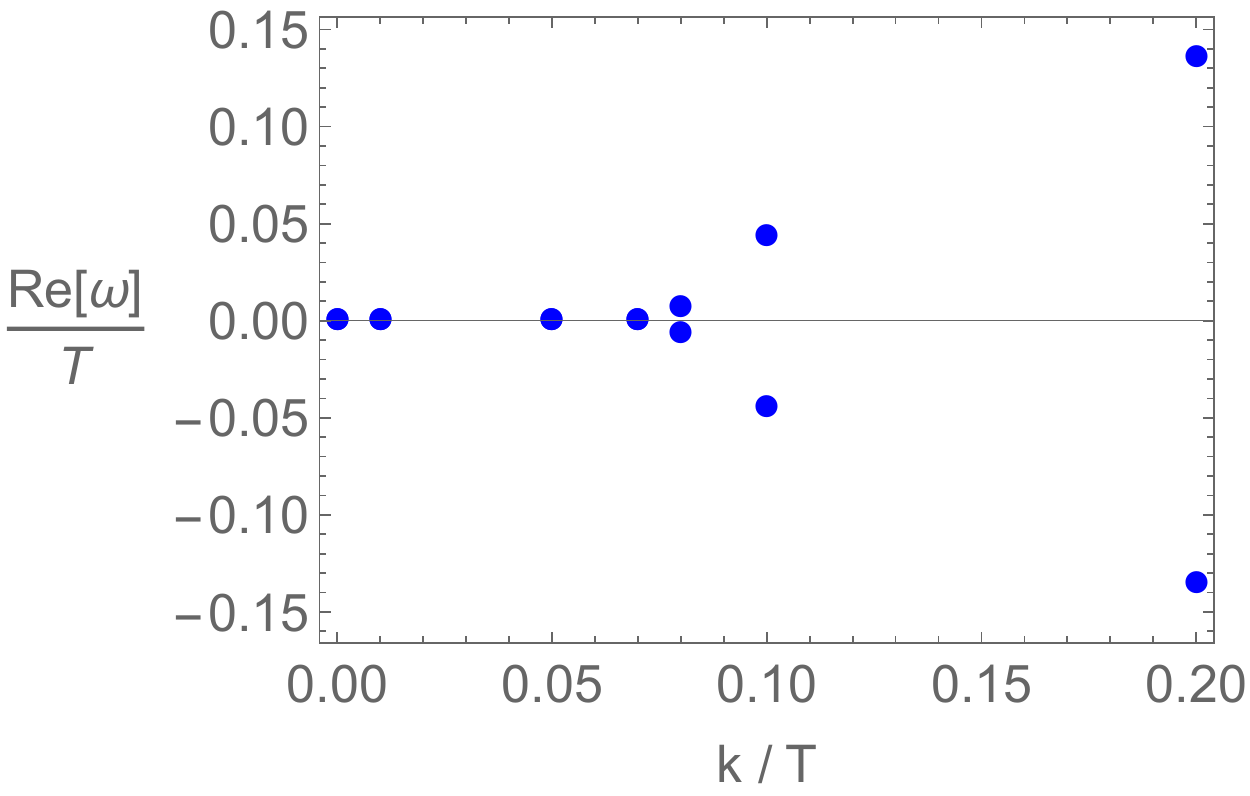} \label{}}
     
     \subfigure[Im($\omega$) at $H/T^2=0$]
     {\includegraphics[width=4.83cm]{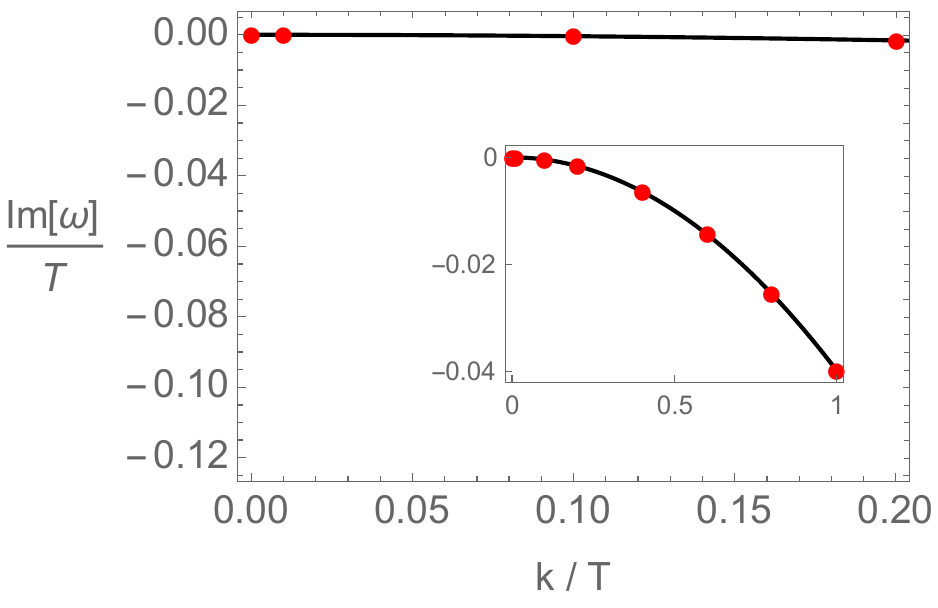} \label{}}
     \subfigure[Im($\omega$) at $H/T^2=3$]
     {\includegraphics[width=4.83cm]{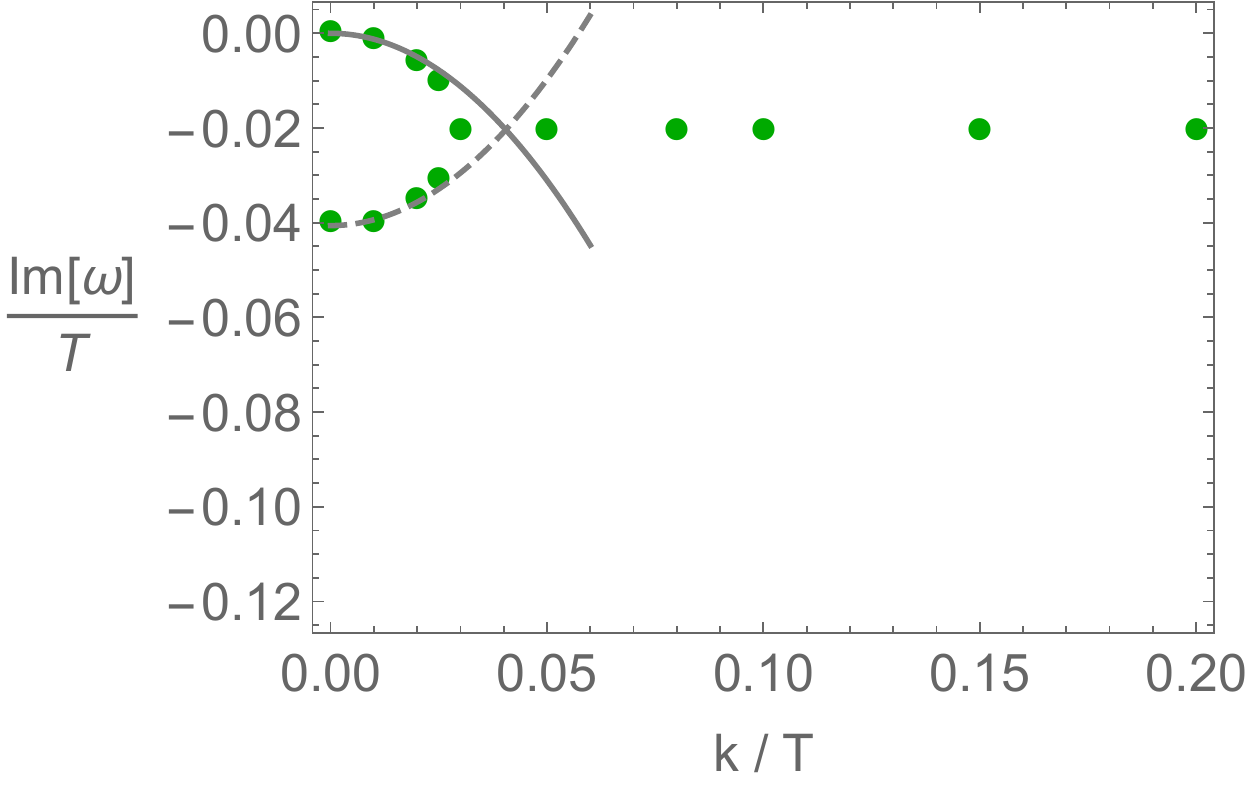} \label{}}
     \subfigure[Im($\omega$) at $H/T^2=5$]
     {\includegraphics[width=4.83cm]{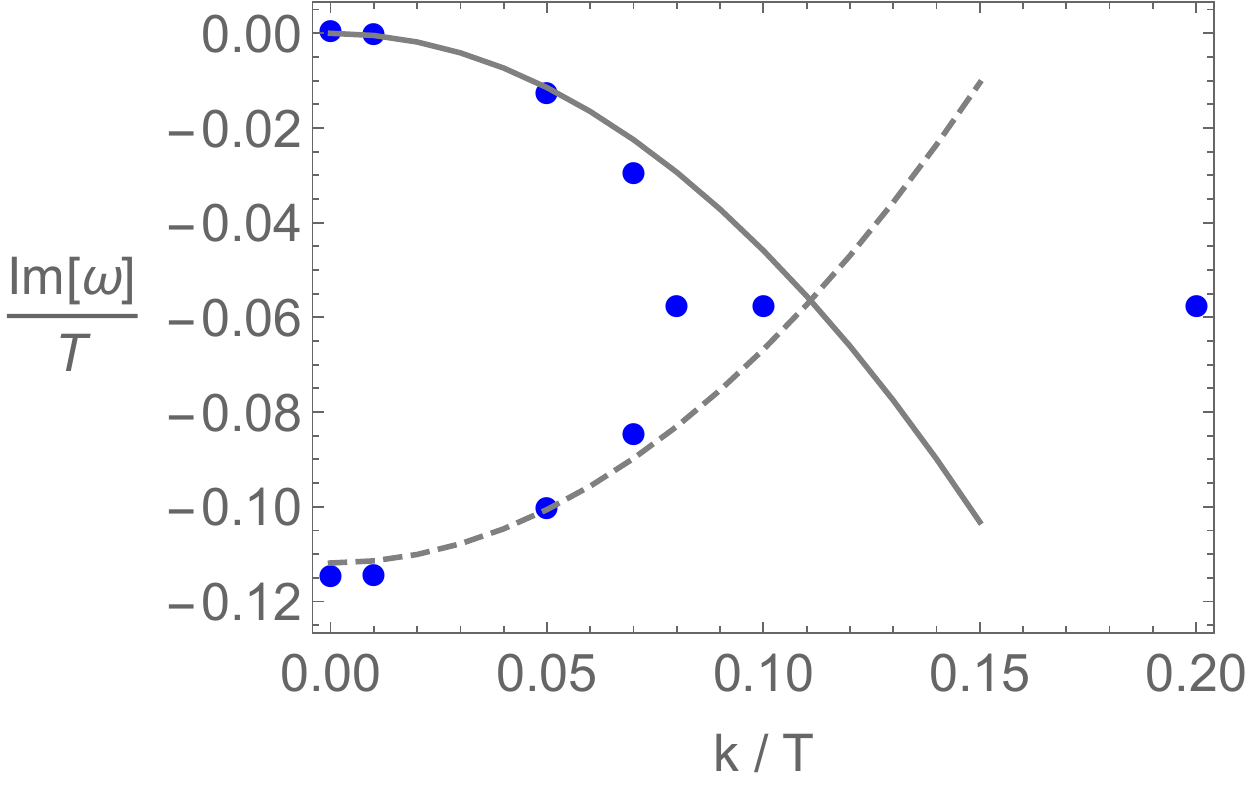} \label{}}
 \caption{Quasi-normal modes vs the magneto-hydrodynamic prediction of the dispersion relations. \textbf{Left:} (a) and (d) are QNMs at $H/T^2=0$. \textbf{Center:} (b) and (e) are QNMs at $H/T^2=3$. \textbf{Right:} (c) and (f) are QNMs at $H/T^2=5$. \textbf{All figures:} colored dots are numerically computed quasi-normal modes and the black solid line is \eqref{SS1}, the gray solid line is \eqref{SD2}, and the gray dashed line is \eqref{SD1}. }\label{PUREEXBFIG}
\end{figure}
The colored dots in all subfigures correspond to the numerically computed quasi-normal modes and the solid or dashed lines are the dispersion relations predicted from magneto-hydrodynamics.
First, when $H=0$ (see (a) and (d)), one can see that the red dots are well matched to the black solid line \eqref{SS1}. Note that the inset of (d) shows that they are still well matched in the large wave vector regime.

\begin{figure}[]
\centering
     \subfigure[Im($\omega$) at $H/T^2=10$]
     {\includegraphics[width=7.2cm]{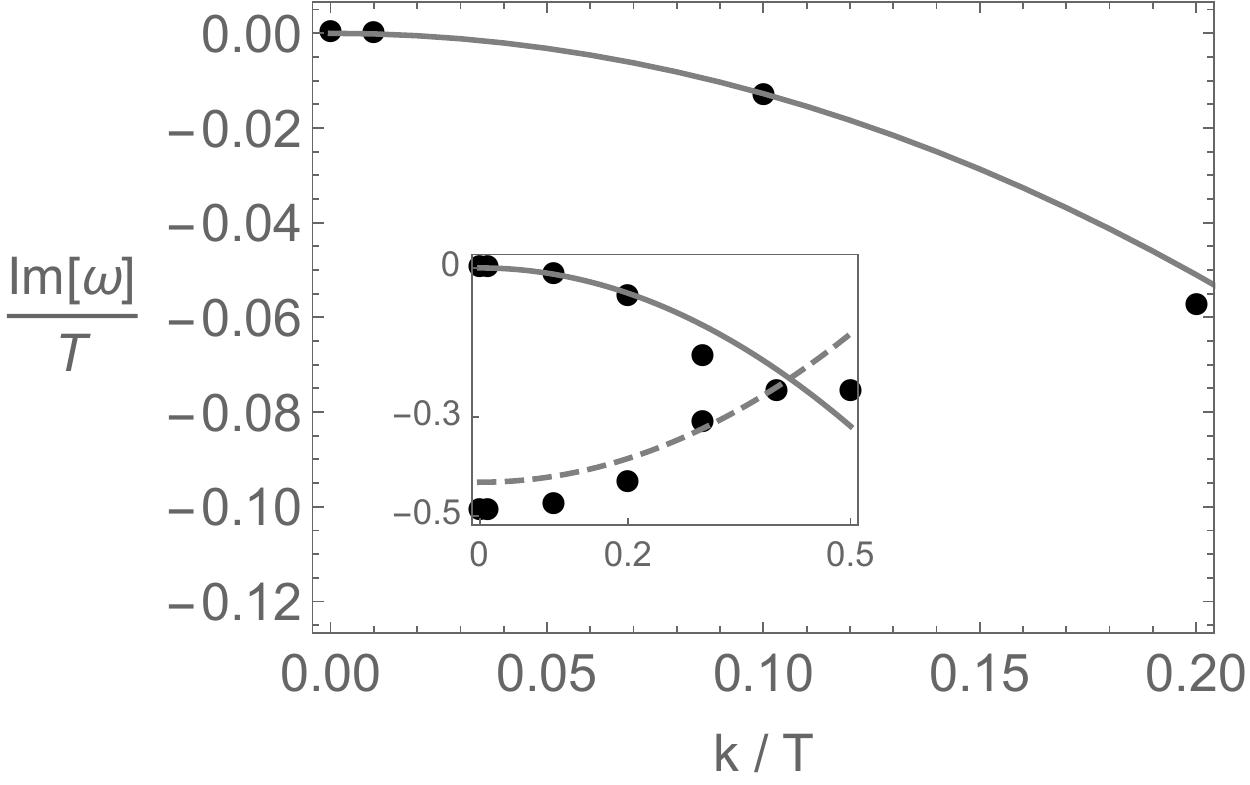} \label{}}
     \subfigure[Im($\omega$) at $H/T^2=50$]
     {\includegraphics[width=7.2cm]{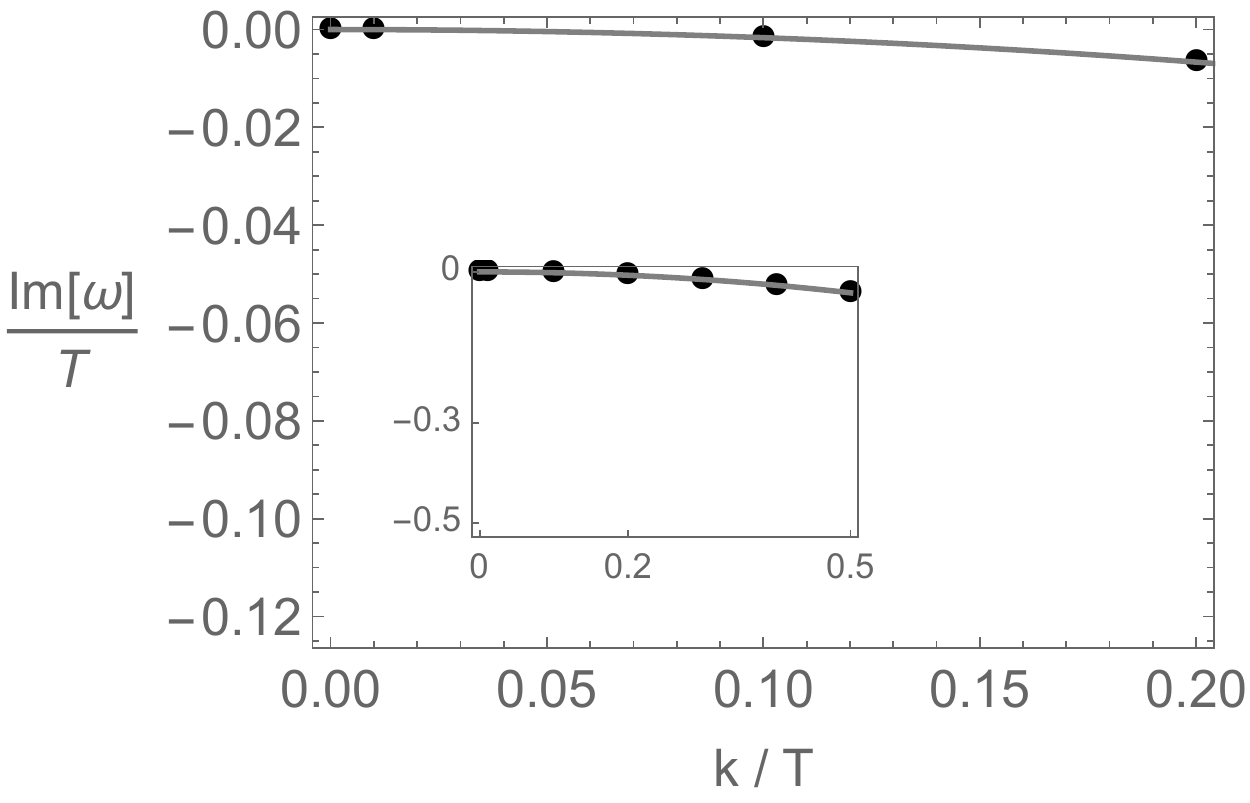} \label{}}
 \caption{Black dots are numerically computed quasi-normal modes and the gray solid line is \eqref{SD2}, the gray dashed line is \eqref{SD1}.  These figures show that even in the large $H$ case, the diffusive mode (gray solid line) could still be a good approximation to the quasi-normal modes, unlike the pseudo diffusive mode  (gray dashed line). The insets display quasi-normal modes in a wider window. Note that a dashed line in the right figure is out of the window.}\label{NEWPLOTS}
\end{figure}

Next, for the finite $H$ case (green or blue dots), quasi-normal modes spectrum follow two diffusive modes from magneto-hydrodynamics: \eqref{SD2} (gray solid line), \eqref{SD1} (gray dashed line).
Note that the dashed line is growing as $k$ increases, so one can notice that the quadratic order in \eqref{SD1} gives positive contribution.
Note also that we can see the deviation for gray dashed line in (f) even near $k/T \sim 0$, i.e., unlike the diffusive mode (gray solid line), the pseudo diffusive mode (gray dashed line) starts to deviate from the magneto-hydrodynamic prediction as we increase $H$ even in the small wave vector regime.
This observation can be more clearly seen for larger value of $H$ in Fig. \ref{NEWPLOTS}.

From the observation above, we may summarize the effect of $H$ as follows:
\begin{itemize}
\item{As $H$ increases, a k-gap\footnote{The gap in the wave vector direction is called k-gap~\cite{Baggioli:2019jcm,Baggioli:2018vfc, Baggioli:2018nnp,Baggioli:2019aqf,Baggioli:2019sio,Hofman:2017vwr,Grozdanov:2018ewh}.} is generated in Re($\omega$): see from (a) to (c) in Fig. \ref{PUREEXBFIG}.}
\item{At finite $H$, two diffusive modes (gray solid line, gray dashed line) generate the dome in the ($k$, Im($\omega$)) plane. As $H$ decreases, the size of the dome vanishes and the quasi-normal mode approaches the damped sound mode: see from (f) to (d) in Fig. \ref{PUREEXBFIG}.}
\item{With strong magnetic fields, the diffusive mode (gray solid line) \eqref{SD2} could be a good approximation to the quasi-normal mode spectrum, unlike the pseudo diffusive mode  (gray dashed line) \eqref{SD1}. Moreover, as we increase $H$, \eqref{SD2} well matches with quasi-normal modes in the larger wave vector regime.}
\end{itemize}
%

%
\subsection{Many-body chaos at $H/T^2 \gg 1$}

Based on the discussion in the previous subsection, let us summarize the hydrodynamic modes\footnote{The hydrodynamic mode is the one such that the frequency goes to zero at zero wave vector.} of the holographic model \eqref{ACTION} as 
\begin{align}\label{}
(H=0): \quad &\omega = \pm \sqrt{\frac{\partial P}{\partial \epsilon}} k \,-\, i  \frac{\eta}{2(\epsilon+P)} \,k^2 \,, \\ 
(H\neq0): \quad &\omega  \,=\, -i \,D_{MHD}\, k^2 \,, \label{MHDHYDRO2} 
\end{align} 
where \eqref{MHDHYDRO2} is valid for any value of $H$.

\paragraph{Pole-skipping phenomena with strong magnetic fields:}
Now let us study the relationship between magneto-hydrodynamics and the pole-skipping point.
One may expect that the pole-skipping point \eqref{PSP} passes though \eqref{MHDHYDRO2} at strong magnetic fields because \eqref{MHDHYDRO2} could be a good approximation to the quasi-normal mode spectrum even with strong magnetic fields.  

\begin{figure}[]
\centering
     \subfigure[$H/T^2=10$]
     {\includegraphics[width=4.83cm]{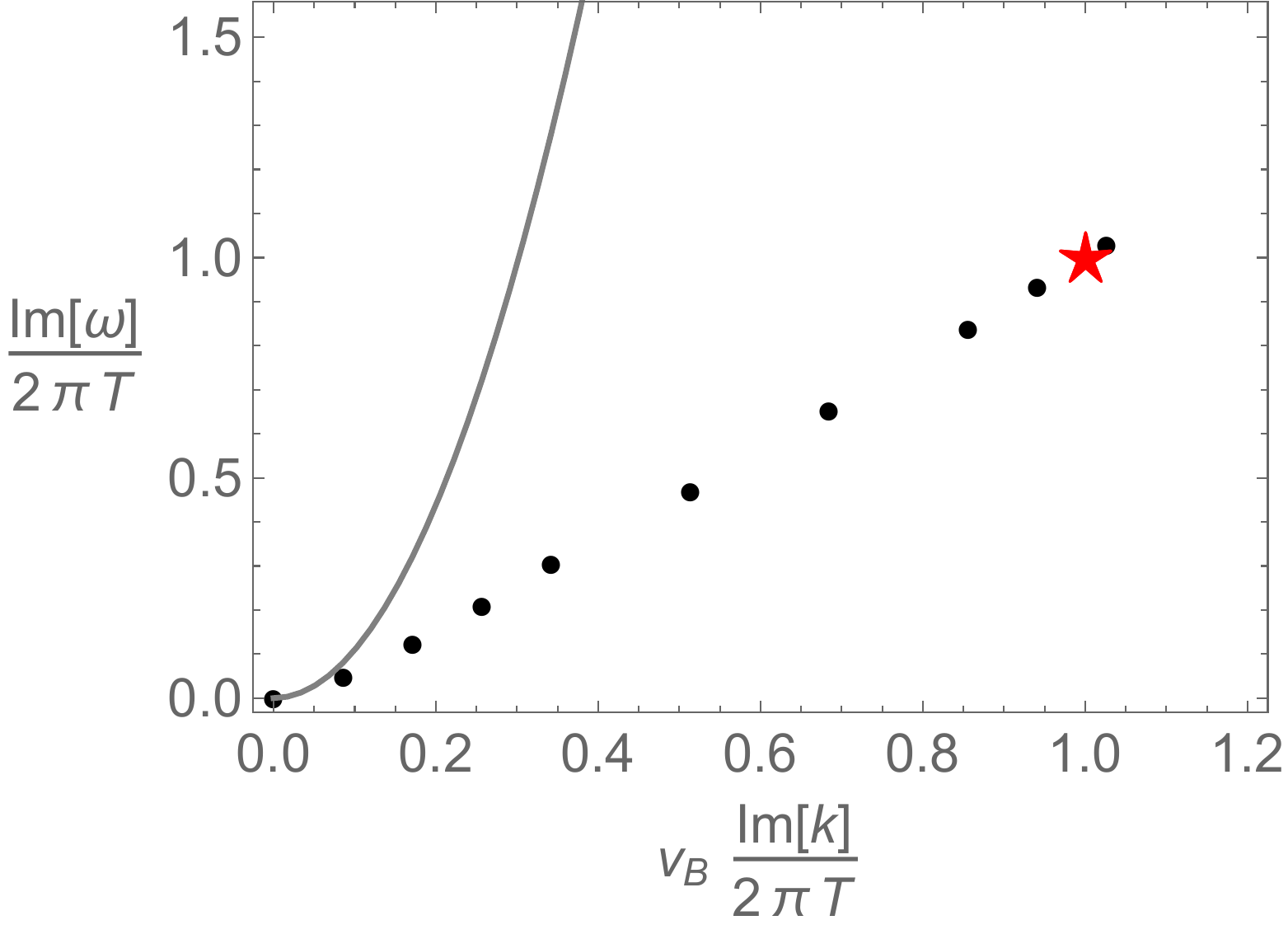} \label{}}
     \subfigure[$H/T^2=100$]
     {\includegraphics[width=4.83cm]{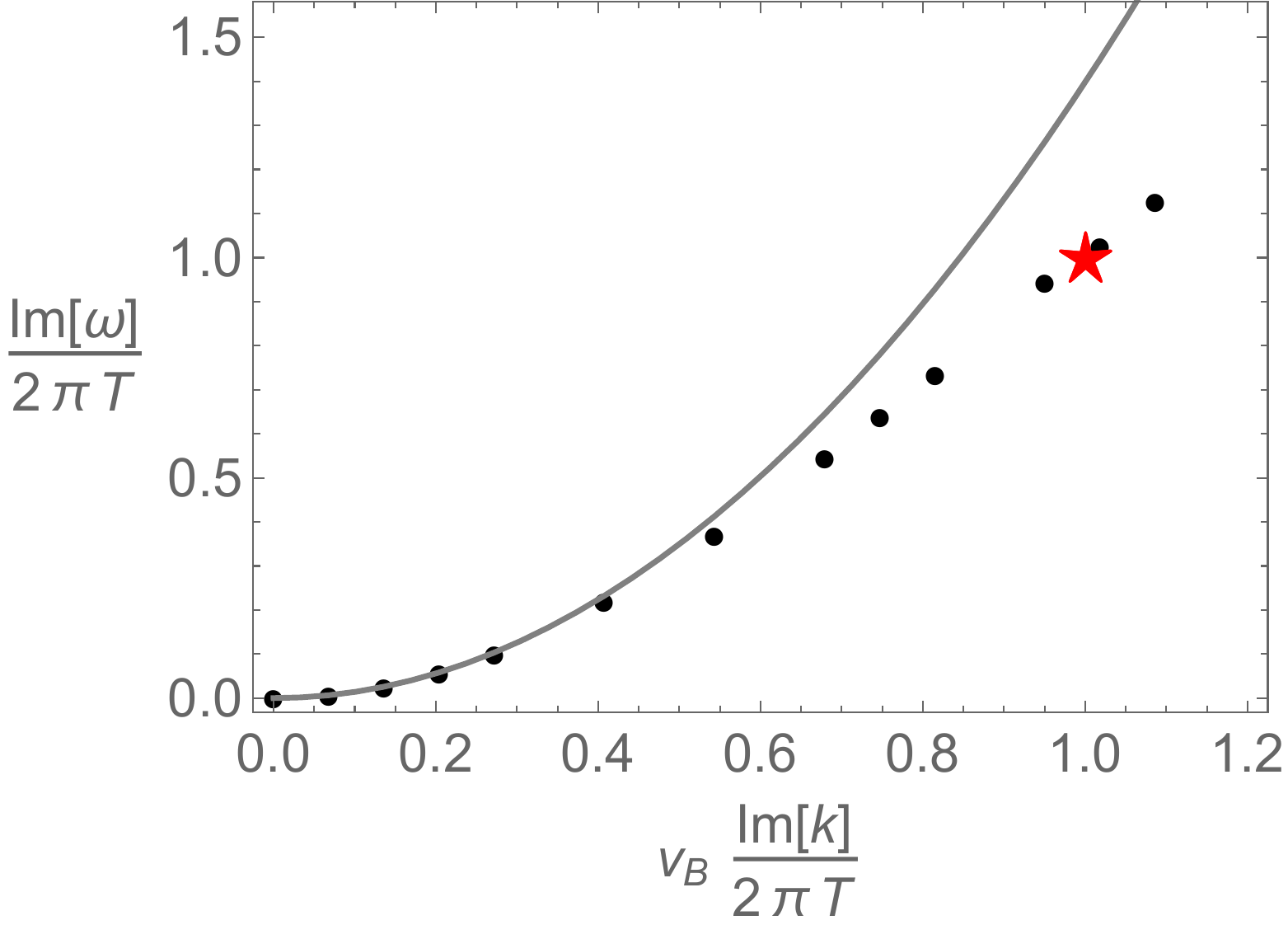} \label{}}
     \subfigure[$H/T^2=10000$]
     {\includegraphics[width=4.83cm]{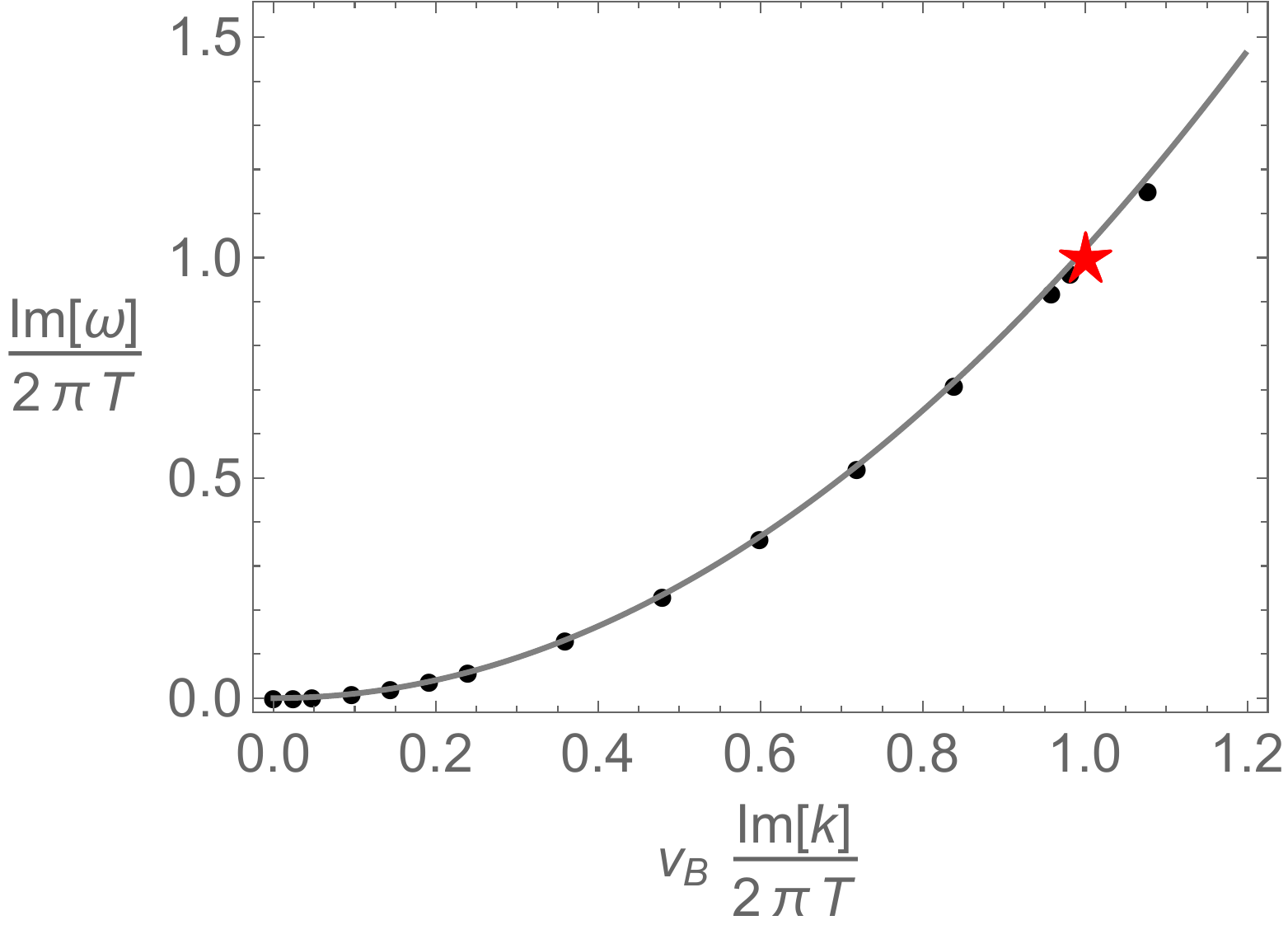} \label{}}
 \caption{Pole-skipping vs magneto-hydrodynamics with various $H$. In all figures, black dots are numerically computed quasi-normal modes, the red star corresponds to the pole-skipping point \eqref{PSP}, and the gray solid line is the diffusive mode \eqref{MHDHYDRO2}.}\label{PSPMHD22}
\end{figure}

We show this expectation in Fig. \ref{PSPMHD22}, with three things for different value of $H$: i) numerically computed quasi-normal modes (black dots); ii) the pole-skipping point \eqref{PSP} (red star); iii) the diffusive mode \eqref{MHDHYDRO2} (gray solid line).

From the figure, at $H/T^2\gg1$, we see that the quasi-normal modes follow a gray solid line \eqref{MHDHYDRO2} in the large wave vector regime and the pole-skipping point (the red star) passes through it.
Thus plugging the pole-skipping point \eqref{PSP} into \eqref{MHDHYDRO2}, we have the following relation at the low temperature limit:
\begin{align}\label{DHOHOH}
\begin{split}
\omega_{*} = - i \, D_{MHD} \, k_{*}^2  \quad\rightarrow\quad D_{MHD} \frac{\lambda_{L}}{v_{B}^2} = 1\,.
\end{split}
\end{align} 

\paragraph{Comparison with the energy diffusion:}
Because of the strong analogy between  \eqref{DHOHOH} and \eqref{EDB22}, one may wonder if $D_{MHD} = D_{E}$. We can mathematically check it in two steps.
First, $D_{MHD}$ is expressed as
\begin{align}\label{COMP1}
\begin{split}
D_{MHD} & := \frac{\partial P}{\partial \epsilon} \frac{\epsilon + P}{\sigma_{Q} H^2} \,=\, \frac{3 r_{h}^3}{2 H^2} + \frac{3}{8 r_{h}}    \,, 
\end{split}
\end{align} 
where we have used \eqref{SMARR}, \eqref{EQSET3}.
Second, $D_{E}$ is given by~\cite{Blake:2015hxa,Li:2019bgc}
\begin{align}\label{}
\begin{split}
D_{E} &:= \frac{\kappa}{c_{\rho}} \,, \quad \kappa = \frac{s^2 \, T}{H^2} \,, \quad c_{\rho} := T\frac{\partial s}{\partial T}  \,, 
\end{split}
\end{align} 
where $\kappa$ is the thermal conductivity and $c_{\rho}$ is the specific heat.
Moreover, using\footnote{We correct the typo in equation (A5) of \cite{Li:2019bgc}.}
\begin{align}\label{}
\begin{split}
\frac{\partial s}{\partial T} \,=\, (8\pi r_{h}) \frac{\partial r_{h}}{\partial T}   \,,  \quad \frac{\partial r_{h}}{\partial T} \,=\, \frac{16\pi r_{h}^4}{12 r_{h}^4 + 3 H^2} \,,
\end{split}
\end{align} 
we can express the energy diffusion constant as
\begin{align}\label{COMP2}
\begin{split}
D_{E} \,=\,  \frac{3 r_{h}^3}{2 H^2} + \frac{3}{8 r_{h}}  \,, 
\end{split}
\end{align} 
which is consistent with \cite{Li:2019bgc}.  Comparing \eqref{COMP1} with \eqref{COMP2}, we can see that ${D}_{MHD}$ is the same as the energy diffusion constant $D_{E}$\footnote{In \cite{Li:2019bgc}, they argued two things: i) the magneto-hydrodynamic diffusion constant is the same as the energy diffusion constant; ii) \eqref{DHOHOH} is satisfied at some specific value of $H/T^2 \sim 33.67$.
The second result is contradictory to our result in that, in our computation, \eqref{DHOHOH} appears at $H/T^2 \gg1$.
One possible answer to this discrepancy is the difference of the setup. Their model is AdS$_{5}$ with anisotropic metric ansatz and their hydrodynamics is built on the existence of the ``transverse" thermal diffusion constant which cannot be seen from our model.
The other possible answer might be related to what they mentioned in their paper: they did not performed a quasi-normal mode computation. So they did not confirm if the pole-skipping point lies on the quasi-normal mode whose low wave vector limit corresponds to the hydrodynamic diffusive mode.
}.

\begin{figure}[]
\centering
     {\includegraphics[width=7.2cm]{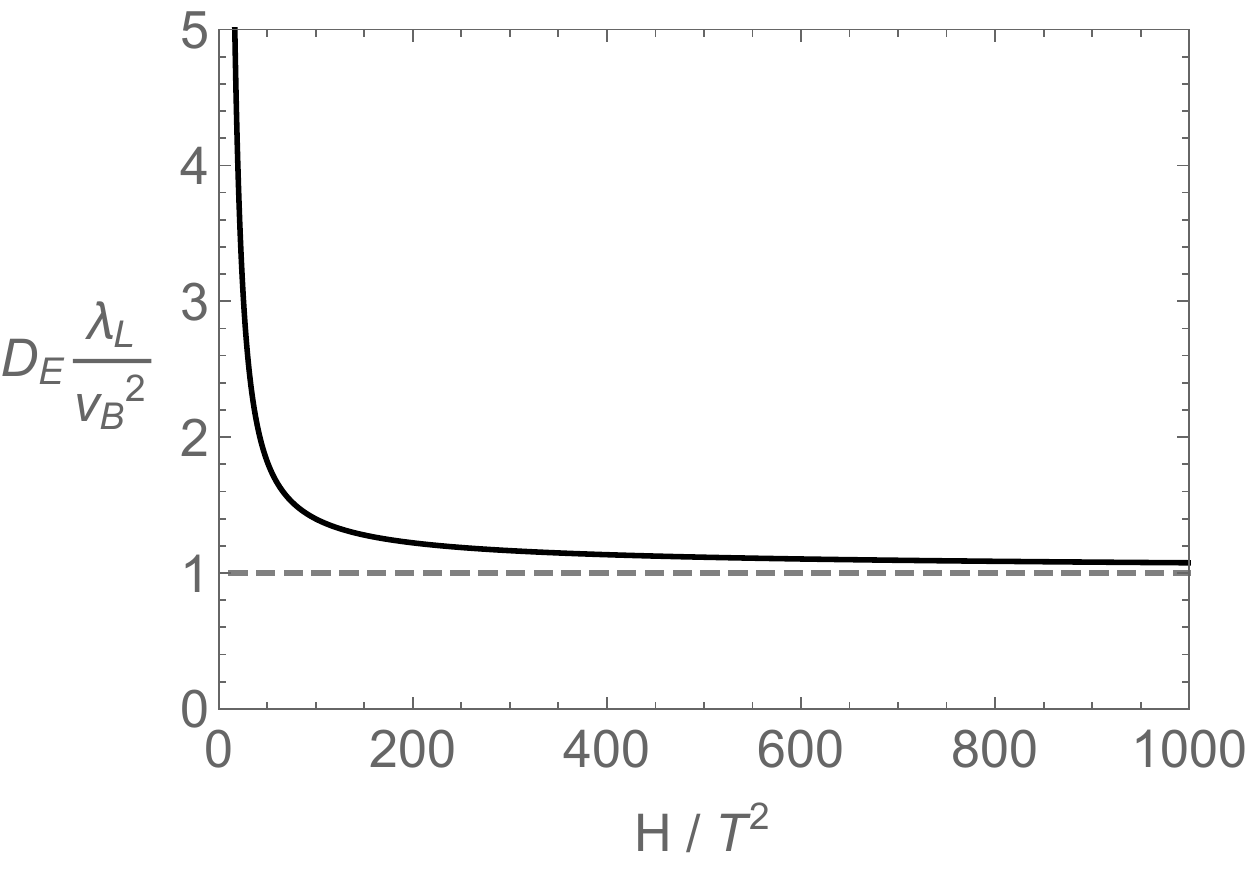} \label{}}
 \caption{The diffusion bounds of $D_{E}$ \eqref{COMP2} with magnetic fields. $\lambda_{L}=2\pi T$, the butterfly velocity $v_{B}$ is \eqref{VTFOMU22}. The dashed line denotes the lower bound \eqref{DHOHOH22}.}\label{DHOHOH22gf}
\end{figure}

Therefore, \eqref{DHOHOH} obtained from the pole-skipping points corresponds to the energy diffusion bound in the presence of magnetic fields:
\begin{align}\label{DHOHOH22}
\begin{split}
D_{E} \frac{\lambda_{L}}{v_{B}^2} = 1\,.
\end{split}
\end{align} 
Note that, as in the axion model in Fig. \ref{LBN1FIG}, the energy diffusion constant with magnetic fields is bounded from below, i.e., \eqref{DHOHOH22} is the lower bound, see Fig. \ref{DHOHOH22gf},  
so that our result might be regarded as the consistency check between pole-skipping analysis for the energy diffusion bound and IR geometry analysis~\cite{Blake:2017qgd} at finite magnetic fields.

One can also check this lower bound \eqref{DHOHOH22} analytically by plugging the zero temperature condition ($H^2=12\,r_{h}^4$) obtained from \eqref{HAWKINGT} into \eqref{COMP2}:
\begin{align}\label{}
\begin{split}
D_{E} = \frac{1}{2\,r_{h}} \,,
\end{split}
\end{align} 
which will produce \eqref{DHOHOH22} with $\lambda_{L}/v_{B}^2 = 2\,r_{h}$\footnote{Recall that $\lambda_{L}=2\pi T$, $v_{B}^2=\pi T/r_{h}$ \eqref{VTFOMU22}.}.


%
%
%
%
%

%
\section{Conclusion}
We have studied the properties of pole-skipping in the sound channel with the breaking of translational invariance.
In particular, our work aims to investigate the pole-skipping phenomenon from the perspective of symmetry breaking patterns: explicit breaking or spontaneous breaking.
For this purpose, we analyze, in detail, not only the axion models, but also the magnetically charged black holes with several methods: the near-horizon analysis, quasi-normal mode computations, and hydrodynamics.

Extending the near-horizon analysis to the case with 
i) a general metric ansatz; 
ii) different symmetry breaking patterns; 
iii) finite magnetic fields, 
we present the followings.
\begin{itemize}
\item{The identity \eqref{IDENTI2} from the $vv$ component of Einstein's equation still holds.}
\item{The form of pole-skipping point \eqref{PSP} does not change.}
\end{itemize}
Moreover, we also found that the near-horizon analysis still can be an independent method to compute the butterfly velocity, which is consistent with the shock wave analysis.
 
In addition to the near-horizon analysis, we also perform the quasi-normal mode computation and show that the quasi-normal mode spectrum is well matched with the following hydrodynamic dispersion relations
\begin{align}
 (EXB): \quad &  \omega = -i \, D_{E} \, k^2 \,, \\
 (SSB): \quad &  \omega = -i \, D_{\phi} \, k^2 \,, \qquad  \omega = \pm \, v_{L} \, k \,-\, i \, D_{L} \, k^2  \label{}\,,
\end{align} 
where the energy diffusion mode in the EXB is for both the axion model and the magnetically charged black hole.

As the temperature is lowered ($m/T\gg1$ or $H/T^2\gg1$), these hydrodynamic dispersions could be a good approximation for quasi-normal modes even out of the hydrodynamic regime, which results in the diffusion lower bound with the pole-skipping point \eqref{PSP} as 
\begin{align}\label{}
\begin{split}
\omega_{*} = - i \, D \, k_{*}^2  \quad\rightarrow\quad D \, \frac{\lambda_{L}}{v_{B}^2} \,=\, 1\,, 
\end{split}
\end{align} 
where $D=D_{E}$ for the EXB and $D=D_{\phi}$ for the SSB.
Therefore, our work supports that the pole-skipping analysis might be a consistent method with the IR geometry analysis to capture the diffusion lower bounds, independent of the symmetry breaking patterns.

%
%

Inspired from this work, it would be interesting to investigate if the pole-skipping points of other sectors of gravitational perturbations are also independent on the symmetry breaking patterns.
Another interesting future direction would be to investigate if it is possible to show the universal bounds of the general IR geometry from the pole-skipping analysis.
For the general IR geometries, the universal diffusion bound with the energy diffusion constant is 
\begin{align}\label{}
\begin{split}
D_{E} \, \frac{\lambda_{L}}{v_{B}^2} \,=\, \frac{z}{2(z-1)}\,, 
\end{split}
\end{align} 
where $z$ is the critical dynamical exponent. 
We hope to address these questions in the near future.

\acknowledgments

We would like to thank  {Navid Abbasi, Andrea Amoretti, Daniel K. Brattan, Richard A. Davison, Karunava Sil}  for valuable discussions and correspondence.  This work was supported by the National Key R$\&$D Program of China (Grant No. 2018FYA0305800), Project 12035016 supported by National Natural Science Foundation of China, the Strategic Priority Research Program of Chinese Academy of Sciences, Grant No. XDB28000000, Basic Science Research Program through the National Research Foundation of Korea (NRF) funded by the Ministry of Science, ICT $\&$ Future Planning (NRF- 2021R1A2C1006791) and GIST Research Institute(GRI) grant funded by the GIST in 2021.

\appendix
%

\section{Magneto-hydrodynamics}\label{appendixa}
Let us review and recap the main features of the magneto hydrodynamics in (2+1) dimensions. We refer the reader to \cite{Buchbinder:2008dc,Buchbinder:2009aa} for more detailed explanation and derivations.
Although we focus on the sound mode coupled to the gravitational sound mode for the study of pole-skipping phenomena in this paper, we also present the remaining shear channel in hydrodynamics for the completeness.

The relevant field theory equation of motions are the conservation  laws:
\begin{align}\label{CONSERVATIONEQ}
\begin{split}
\partial^{\nu} T_{\mu\nu} = F_{\mu\nu} J^{\nu} \,, \qquad
\partial_{\mu} J^{\mu} = 0,
\end{split}
\end{align} 
where $T_{\mu\nu}$ is the stress energy tenser, $J^{\mu}$ is the current and $F_{\mu\nu}$ is the external electromagnetic field.
For the case under the consideration, $F_{\mu\nu}$ is taken to be magnetic as
\begin{align}\label{}
\begin{split}
F_{tx} = 0\,, \quad F_{ty} = 0\,, \quad F_{ij} = \epsilon_{ij} H\,,
\end{split}
\end{align} 
where $i,j = (x, y)$.
To first order in derivatives, $T_{\mu\nu}$ and $J^{\mu}$ can be given by the standard expression:
\begin{align}\label{TMUNUEQ}
\begin{split}
T^{\mu\nu} = \epsilon u^{\mu} u^{\nu} + P \Delta^{\mu\nu} - \eta\left(\Delta^{\mu\alpha} \Delta^{\nu\beta}(\partial_{\alpha} u_{\beta} + \partial_{\beta} u_{\alpha})  - \Delta^{\mu\nu} \partial_{\gamma}u^{\gamma} \right) \,,
\end{split}
\end{align} 
where $\Delta^{\mu\nu} = \eta^{\mu\nu} + u^{\mu} u^{\nu}$, $u^{\mu}$ is the fluid 3-velocity, $\epsilon$ and $P$ are the energy density and the pressure respectively, and $\eta$ is the shear viscosity.
Similarly, the current is given by
\begin{align}\label{JCURRENTEQ}
\begin{split}
J^{\mu} = \rho u^{\mu} + \sigma_{Q} \Delta^{\mu\nu} (-\partial_{\nu} \mu + F_{\nu\alpha} u^{\alpha} + \frac{\mu}{T}\partial_{\nu} T) \,,
\end{split}
\end{align} 
where $\rho$ is the charge density, $\mu$ is the chemical potential, $T$ is the temperature and $\sigma_{Q}$ is the conductivity coefficient.
For the study of fluctuation around the equilibrium in which
\begin{align}\label{}
\begin{split}
u^{\mu} = (1, 0, 0) \,, \qquad T = \text{constant}\,, \qquad \mu = \text{constant} \,,
\end{split}
\end{align} 
we choose $(\delta u_{x}, \, \delta u_{y},\, \delta T,\, \delta \mu)$ as the independent variables. Then, with the plane wave form $e^{-i \omega t + i k x}$, one can find the eight relevant fluctuations: 
\begin{align}\label{FLUCFLUC}
\begin{split}
\delta T^{tt}\,,\,\,  \delta T^{tx}\,, \,\, \delta T^{ty}\,,\,\,  \delta T^{xy}\,, \,\, \delta T^{xx}\,, \,\, 
\delta J^{t}\,, \,\, \delta J^{x}\,, \,\, \delta J^{y}\,.
\end{split}
\end{align} 
For the specific form of \eqref{FLUCFLUC}, see equation (2.14) and (2.15) in \cite{Buchbinder:2008dc}. 
After putting all the fluctuations \eqref{FLUCFLUC} into the equation \eqref{CONSERVATIONEQ} and performing a Fourier transformation, we can get the four coupled equations:
\begin{align}\label{}
\begin{split}
 0 &= \omega \left(  \left(\frac{\partial{\epsilon}}{\partial{\mu}}\right)_{T}  \delta \mu +   \left(\frac{\partial{\epsilon}}{\partial{T}}\right)_{\mu} \delta T    \right) - k (\epsilon + P) \delta u_{x} \,, \\
 0 &= \omega (\epsilon + P) \delta u_{x} - k \left(  \left(\frac{\partial{P}}{\partial{\mu}}\right)_{T}  \delta \mu +   \left(\frac{\partial{P}}{\partial{T}}\right)_{\mu} \delta T    \right) + i k^2 \eta \, \delta u_{x} + i \sigma_{Q} H^2 \delta u_{x} + i H \rho \delta u_{y} \,, \\
 0 &= \omega (\epsilon + P) \delta u_{y} - k H \sigma_{Q} \left(\delta \mu - \frac{\mu}{T} \delta T\right) - i H \rho \delta u_{x} + i \sigma_{Q} H^2 \delta u_{y} + i k^2 \eta \delta u_{y} \,, \\
 0 &= \omega \left(  \left(\frac{\partial{\rho}}{\partial{\mu}}\right)_{T}  \delta \mu +   \left(\frac{\partial{\rho}}{\partial{T}}\right)_{\mu} \delta T    \right) - k \rho \delta u_{x} + k \sigma_{Q} H \delta u_{y} + i k^2 \sigma_{Q} \left(\delta\mu - \frac{\mu}{T}\delta T \right) \,.
\end{split}
\end{align} 
Although these equations are all coupled each other, if we consider the zero charge density($\rho=0$) and no chemical potential($\mu=0$) which is the same condition\footnote{In addition to zero charge condition, motivated by M2-brane magneto hydrodynamics, we may set $\left(\frac{\partial \rho}{\partial T}\right)_{\mu} = \left(\frac{\partial \epsilon}{\partial \mu}\right)_{T} = 0$, $\left(\frac{\partial \rho}{\partial \mu}\right)_{T} \neq 0$, $\left(\frac{\partial \epsilon}{\partial T}\right)_{\mu} \neq 0$.} used for the holographic computation in this paper, these equations can be simplified and decoupled into two independent pairs: i) sound channel; ii) shear channel.

\paragraph{Sound channel:}
The first pair of equation is called sound channel and it reads as
\begin{align}\label{SOUNDCHNNEL}
\begin{split}
&\omega \left(\frac{\partial\epsilon}{\partial T}\right)_{\mu} \delta T - k(\epsilon + P) \delta u_{x} = 0\,, \\
& \omega(\epsilon + P)\delta u_{x} - k\left(\frac{\partial P}{\partial T}\right)_{\mu} \delta T + i k^2 \eta \delta u_{x} + i \sigma_{Q} H^2 \delta u_{x} = 0 \,.
\end{split}
\end{align} 
Combining these two equations, we can obtain one equation about $\omega$, which is the second order equation for $\omega$. Then, by solving it, we get two dispersion relations $\omega(k)$.
When $H=0$, the equation gives the sound mode with the dispersion relation\footnote{We correct the typo in \cite{Buchbinder:2008dc,Buchbinder:2009aa}: there should be factor $1/2$ in $k^2$ order.}:
\begin{align}\label{SCDSP1}
\begin{split}
(H=0): \quad \omega = \pm \sqrt{\frac{\partial P}{\partial \epsilon}} k \,-\, i  \frac{\eta}{2(\epsilon+P)} \,k^2  \,+\, \mathcal{O}(k^3)\,.
\end{split}
\end{align} 
On the other hand, when the magnetic field is turned on, the equation gives the dramatically changed dispersion relations:
\begin{align}\label{SCDSP2}
\begin{split}
(H\neq0): \quad \omega = -i \frac{\sigma_{Q} H^2}{\epsilon+P} + \mathcal{O}(k^2) \,, \qquad
\omega = -i  \frac{\partial P}{\partial \epsilon} \frac{\epsilon + P}{\sigma_{Q} H^2} \,k^2 + \mathcal{O}(k^4) \,.
\end{split}
\end{align} 
As it has been explained in \cite{Buchbinder:2008dc,Buchbinder:2009aa}, this drastic change occurs because, as one can see from \eqref{SOUNDCHNNEL}, the small $H$ limit does not commute with the hydrodynamic limit of small $\omega$ and $k$. Thus, the magnetic field cannot be considered as a small perturbation.

\paragraph{Shear channel:}
The second pair of equation is called shear channel and it gives
\begin{align}\label{SHEARCHNNEL}
\begin{split}
&\omega(\epsilon+P)\delta u_{y} - k H \sigma_{Q} \delta {\mu} + i \sigma_{Q} H^2 \delta u_{y} + i k^2 \eta \delta u_{y} = 0 \,, \\
&\omega \left( \frac{\partial \rho}{\partial \mu} \right)_{T} \delta \mu + k \sigma_{Q} H \delta u_{y} + i k^2 \sigma_{Q} \delta \mu = 0 \,.
\end{split}
\end{align} 
When $H=0$, these equation gives two diffusive modes:
\begin{align}\label{MAGNETO1}
\begin{split}
(H=0): \quad \omega = -i  \frac{\sigma_{Q}}{\left(\frac{\partial\rho}{\partial\mu}\right)_{T}} \,k^2  \,, \qquad
\omega = -i \frac{\eta}{\epsilon+P} k^2 \,,
\end{split}
\end{align} 
where the first (second) one is called charge (shear) diffusion mode. 
In the presence of magnetic field, these diffusive mode also undergo the dramatic change:
\begin{align}\label{MAGNETO2}
\begin{split}
(H\neq0): \quad \omega = -i \frac{\sigma_{Q} H^2}{\epsilon+P} + \mathcal{O}(k^2) \,, \qquad
\omega = -i  \frac{\eta}{H^2 \left( \frac{\partial \rho}{\partial \mu} \right)_{T}} \,k^4 + \mathcal{O}(k^6) \,.
\end{split}
\end{align} 
%

%
\section{The determinant method}\label{appendixb}

In this section, we briefly present how to compute the quasi-normal modes using the determinant method~\cite{Kaminski:2009dh}. We here focus on the sound channel \eqref{FLUCOURSETUP}, but we can also use the method for \eqref{PW2PW2PW2}, see more details in \cite{Blake:2018leo, Davison:2014lua, Ammon:2019apj}.

First, we choose the following deffeomorphism and gauge-invariant combinations~\cite{Buchbinder:2008dc,Buchbinder:2009aa}:
\begin{align}
\begin{split}
Z_{H} &:= \frac{4 k}{\omega} \, h_{t}^{x} \,+\,  2 h_{x}^{x} - \left( 2 - \frac{k^2}{\omega^2}\frac{f'(r)}{r} \right) h_{y}^{y}  + \frac{2k^2}{\omega^2}\frac{f(r)}{r^2} h_{t}^{t}  \,, \\
Z_{A} &:= a_{y}  \,+\, \frac{i H}{2k} \left(h_{x}^{x}-h_{y}^{y}\right) \,. 
\end{split}
\end{align}
where we raised an index on fluctuation fields using the background metric \eqref{bgc}.
Then, we can obtain the gauge invariant second order equations for $Z_{H}$ and $Z_{A}$ of the following form:
\begin{align}\label{ZAZHEOM}
\begin{split}
&0 \,=\, A_{H}\,Z_{H}'' \,+\, B_{H}\,Z_{H}' \,+\, C_{H}\,Z_{H} \,+\, D_{H}\,Z_{A}' \,+\, E_{H}\,Z_{A} \,, \\
&0 \,=\, A_{A}\,Z_{A}'' \,\,+\, B_{A}\,Z_{A}' \,\,+\, C_{A}\,Z_{A} \,\,+\, D_{A}\,Z_{H}' \,\,+\, E_{A}\,Z_{H} \,.
\end{split}
\end{align}
Since the coefficients of equations are lengthy and cumbersome we will not write them in the paper.

Next, we solve the equations of motion \eqref{ZAZHEOM} with two boundary conditions: one from incoming boundary condition at the horizon and the other from the AdS boundary.
First, near the horizon ($r\rightarrow r_{h}$) the variables are expanded as 
\begin{align}\label{APPENHORIZON}
\begin{split}
Z_{H} = (r-r_{h})^{\nu_{\pm}} \left( Z_{H}^{(I)} \,+\, Z_{H}^{(II)} (r-r_{h}) \,+\, \dots   \right ) \,, \\
Z_{A} = (r-r_{h})^{\nu_{\pm}} \left( Z_{A}^{(I)} \,+\, Z_{A}^{(II)} (r-r_{h}) \,+\, \dots   \right ) \,.
\end{split}
\end{align}
where $\nu_{\pm}:= \pm i\omega/4 \pi T$ and we choose $\nu_{-}:= - i\omega/4 \pi T$ for the incoming boundary condition.
After plugging \eqref{APPENHORIZON} into equations \eqref{ZAZHEOM}, one can find that higher horizon coefficients are determined by two independent horizon variables : $(Z_{H}^{(I)}, Z_{A}^{(I)})$.

Near the AdS boundary ($r\rightarrow \infty$), the asymptotic behavior of solutions read
\begin{align}\label{}
\begin{split}
&Z_{H} = Z_{H}^{(S)} \, r^{0} \,(1 \,+\, \dots) \,+\, Z_{H}^{(R)} \, r^{-3} \,(1 \,+\, \dots) \,, \\
&Z_{A} = Z_{A}^{(S)} \, r^{0} \,(1 \,+\, \dots) \,+\, Z_{A}^{(R)}\, r^{-1} \,(1 \,+\, \dots) \,,
\end{split}
\end{align}
where the superscripts mean that $(S)$ is the source term and $(R)$ is a response term. 

Now, we compute the quasi-normal modes by employing the determinant method. Using the shooting method, we can construct the matrix of sources with the source terms near the boundary:
\begin{align}\label{APPENSMATA}
\begin{split}
S = \left(\begin{array}{cc}Z_{H}^{(S)(I)} & Z_{H}^{(S)(II)} \\Z_{A}^{(S)(I)} & Z_{A}^{(S)(II)}\end{array}\right) \,,
\end{split}
\end{align}
where the $S$-matrix is $2\times2$ matrix because we can get two independent solutions with two independent shooting variables at the horizon \eqref{APPENHORIZON}. Note that $I (II)$ in \eqref{APPENSMATA} denotes that the source terms are obtained by the $I (II)$ th shooting.

Finally, the quasi-normal mode spectrum, the dispersion relation in which the holographic Green's functions have a pole, can be given by the value of ($\omega, k$) for which the determinant of $S$-matrix \eqref{APPENSMATA} vanishes.

%
\section{The hydrodynamic cyclotron mode}\label{appendixc}

In \cite{Hartnoll:2007ip,Hartnoll:2007ih,Blake:2015hxa}, investigating the hydrodynamic cyclotron mode with the magneto-transport such as the electric conductivity, they found that the dominant feature of the conductivity is the presence of a cyclotron pole:
\begin{align}\label{PLASMA}
\begin{split}
\omega \,=\, \pm \omega_{c} \,-\, i \gamma \,, 
\end{split}
\end{align} 
which corresponds to damped cyclotron oscillations of the plasma mode.
This mode is composed of two frequencies: i) the cyclotron frequency $\omega_{c}$ proportional to the charge density; ii) the damping frequency $\gamma$.
The damping frequency is given by
\begin{align}\label{GAMMAIMMA}
\begin{split}
\gamma \,=\, \frac{\sigma_{Q} H^2}{\epsilon + P} \,,
\end{split}
\end{align} 
and this corresponds to the contribution to the momentum relaxation arising from the external magnetic field. 

Thus we may rewrite $D_{MHD}$ \eqref{COMP1} with the momentum dissipation rate $\Gamma$ as,
\begin{align}\label{}
\begin{split}
D_{MHD} = \frac{\partial{P}}{\partial{\epsilon}} \Gamma^{-1} \,, \qquad \Gamma:=\gamma\,.
\end{split}
\end{align} 
Note that a momentum dissipation rate (or damping term) can be present even without the external impurities (e.g., without axion fields), and is usually thought of as a consequence of collisions between particles and anti-particles undergoing cyclotron orbits in the opposite directions.
{
Note also that, from the recent development of magneto-transport \cite{Amoretti:2021fch,Amoretti:2020mkp,Amoretti:2019buu}, discussions in this section may have the correction if the magnetic field is no longer taken to be of order one in derivatives. We thank Daniel K. Brattan, Andrea Amoretti for pointing this out.
}


\bibliographystyle{JHEP}
\providecommand{\href}[2]{#2}\begingroup\raggedright\endgroup

\end{document}